\documentclass[a4paper]{article}

\usepackage{times}          
\usepackage[utf8]{inputenc} 
\usepackage[T1]{fontenc}    
\usepackage[english]{babel}
\usepackage{amssymb, amsmath}
\usepackage{url}            
\usepackage{relsize}        
\usepackage{booktabs,arydshln}       
\usepackage{multirow}
\usepackage{amsfonts}       
\usepackage{nicefrac}       
\usepackage{microtype}      
\usepackage{pifont}         
\usepackage[pdftex]{graphicx}
\usepackage{subcaption}
\usepackage[figuresleft]{rotating}
\usepackage[numbers,sort,compress]{natbib}
\usepackage{hyperref}       
\usepackage[accepted]{icml2017}
\usepackage{color}

\icmltitlerunning{Automating the search for a patent's prior art with a full text similarity search}

\begin{document}
\twocolumn[
\icmltitle{Automating the search for a patent's prior art with a full text similarity search}

\icmlsetsymbol{equal}{*}

\begin{icmlauthorlist}
\icmlauthor{Lea Helmers$^1$}{equal}
\icmlauthor{Franziska Horn$^1$}{equal}
\icmlauthor{Franziska Biegler$^2$}{}
\icmlauthor{Tim Oppermann$^2$}{}
\icmlauthor{Klaus-Robert Müller$^{1,3,4}$}{}\\
$^1$Machine Learning Group, Technische Universität Berlin, Berlin, Germany\\
$^2$Pfenning, Meinig \& Partner mbB, Berlin, Germany\\
$^3$Department of Brain and Cognitive Engineering, Korea University, Anam-dong, Seongbuk-gu, Seoul 02841, Korea\\
$^4$Max-Planck-Institut für Informatik, Saarbrücken, Germany\\
  \texttt{franziska.horn@campus.tu-berlin.de, klaus-robert.mueller@tu-berlin.de}
\end{icmlauthorlist}

\icmlkeywords{patents, search, prior art, information retrieval}

\vskip 0.3in
]
\printAffiliationsAndNotice{\icmlEqualContribution}

\begin{abstract}
More than ever, technical inventions are the symbol of our society's advance. Patents guarantee their creators protection against infringement. For an invention being patentable, its novelty and inventiveness have to be assessed. Therefore, a search for published work that describes similar inventions to a given patent application needs to be performed. Currently, this so-called search for prior art is executed with semi-automatically composed keyword queries, which is not only time consuming, but also prone to errors. In particular, errors may systematically arise by the fact that different keywords for the same technical concepts may exist across disciplines.\\
In this paper, a novel approach is proposed, where the full text of a given patent application is compared to existing patents using machine learning and natural language processing techniques to automatically detect inventions that are similar to the one described in the submitted document. Various state-of-the-art approaches for feature extraction and document comparison are evaluated. In addition to that, the quality of the current search process is assessed based on ratings of a domain expert. The evaluation results show that our automated approach, besides accelerating the search process, also improves the search results for prior art with respect to their quality.
\end{abstract}

\section{Introduction}
A patent is the exclusive right to manufacture, use, or sell an invention and is granted by the government's patent offices \cite{wipohandbook}. For a patent to be granted, it is indispensable that the described invention is not known or easily inferred from the so-called prior art, where prior art includes any written or oral publication available before the filing date of the submission. Therefore, for each application that is submitted, the responsible patent office performs a search for related work to check if the subject matter described in the submission is inventive enough to be patentable \cite{wipohandbook}. Before handing in the application to the patent office, the inventors will usually consult a patent attorney, who represents them in obtaining the patent. In order to assess the chances of the patent being granted, the patent attorney often also performs a search for prior art.

When searching for prior art, patent officers and patent attorneys are currently mainly relying on simple keyword searches such as those implemented by the \textsc{Espacenet} tool from the \textit{European Patent Office}, the \textsc{TotalPatent} software developed by \textsc{LexisNexis}, or the \textsc{PatSnap} patent search, all of which provide very limited \emph{semantic} search options. These search engines often fail to return relevant documents and due to constraints regarding the length of the entered search text, it is usually not possible to consider a patent application's entire text for the search, but merely query the database for specific keywords.

Current search approaches for prior art therefore require a significant amount of manual work and time, as given a patent application, the patent officer or attorney has to manually formulate a search query by combining words that should match documents describing similar inventions \cite{Alberts2017book1}. Furthermore, these queries often have to be adapted several times to optimize the output of the search \cite{golestan2015term,Tseng2007}. A main problem here is that regular keyword searches do not inherently take into account synonyms or more abstract terms related to the given query words. This means, if for an important term in the patent application a synonym, such as \emph{wire} instead of \emph{cable}, or a more specialized term, such as \emph{needle} instead of \emph{sharp object}, has been used in an existing document of prior art, a keyword search might fail to reveal this relation unless the alternative term was explicitly included in the search query. This is relevant as it is quite common in patent texts to use very abstract and general terms for describing an invention in order to maximize the protective scope \cite{tannebaum2015patnet,Andersson2017book9}.
A line of research \cite{kando2000workshop,alberts2011introduction,Lupu2017book2,lupu2013patent,shalaby2017patent} has focused on automatically expanding the manually composed queries, e.g., to take into account synonyms collected in a thesaurus \cite{magdy2011study,Lupu2017book2} or include keywords occurring in related patent documents \cite{fujii2007enhancing,mahdabi2012learning,mahdabi2014effect}. Yet, with iteratively augmented queries -- be it by manual or automatic extension of the query -- the search for prior art remains a very time consuming process.

Furthermore, a keyword-based search for prior art, even if done with most professional care, will often produce suboptimal results (as we will see e.g.\ later in this paper and Supporting Information~\ref{S4b}). With possibly imperfect queries, it must be assumed that relevant documents are missed in the search, leading to \textit{false negatives} (FN). On the other hand, query words can also appear in texts that, nonetheless, have quite different topics, which means the search will additionally yield many \textit{false positives} (FP). When searching for prior art for a patent application, the consequences of false positives and false negatives are quite different. While false positives cause additional work for the patent examiner, who has to exclude the irrelevant documents from the report, false negatives may lead to an erroneous grant of a patent, which can have profound legal and financial implications for both the owner of said patent as well as competitors \cite{Trippe2017book5}.

\subsection{An approach to automate the search for prior art}
To overcome some of these disadvantageous aspects of current keyword-based search approaches, it is necessary to decrease the manual work and time required for conducting the search itself, while increasing the quality of the search results by avoiding irrelevant patents from being returned, as well as automatically accounting for synonyms to reduce false negatives. This can be achieved by comparing the patent application with existing publications based on their \emph{entire texts} rather than just searching for specific keywords. By considering the entire texts of the documents, much more information, including the context of keywords used within the respective documents, is taken into account. For humans it is of course infeasible to read the whole text of each possibly relevant document. Instead, state-of-the-art text processing techniques can be used for this task.

\begin{sidewaysfigure*}[p]
\includegraphics[width=\linewidth]{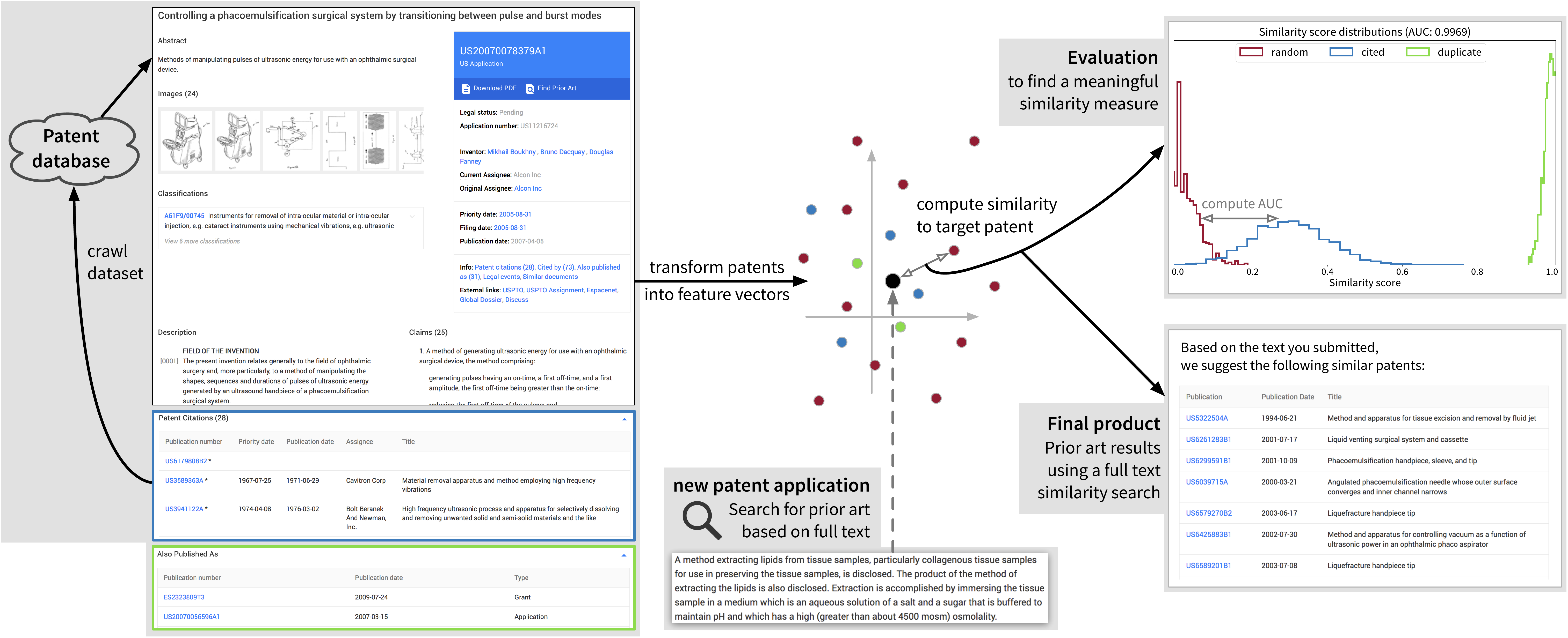}
\caption{{\bf Illustration of the presented novel approach to the search for a patent's prior art.} First, a dataset of patent applications is obtained from a patent database using a few manually selected seed patents and recursively including the patent applications they cite. Then, the patent texts are transformed into feature vectors and the similarity between two documents is computed based on said feature vectors. Finally, patents that are considered as very similar to a new target patent application are returned as possible prior art. An appropriate similarity measure for this process should assign high similarity scores to related patents (e.g.\ where one patent was cited in the search report of the other) and low scores to unrelated (randomly paired) patents. We compare different similarity measures by quantifying the overlap between the respective similarity score distributions of pairs of related documents and randomly paired patents using the AUC score.}\label{fig1}
\end{sidewaysfigure*}

This paper describes a novel approach to automate the search for prior art with \textit{natural language processing} (NLP) and \emph{machine learning} (ML) techniques, such as neural network language models, in order to make it more efficient and accurate. The essence of this idea is illustrated in Fig~\ref{fig1}. We first obtain a dataset of related patents from a patent database by using a few manually selected seed patents and then recursively adding the patents or patent applications that are cited by the documents already included in the dataset. The patent texts are then transformed into numerical feature vectors, based on which the similarity between two documents can be computed. We evaluate different similarity measures by comparing the prior art suggested by our automated approach to those documents that were originally cited in a patent's search report and, in a second step, to documents considered relevant prior art for this patent by a patent attorney.
By analyzing and comparing different approaches for computing full text similarities between patent documents, we aim to identify a similarity measure based on which it is possible to automatically and reliably select relevant prior art given, e.g., the draft of a new patent application.

The remainder of the paper is structured as follows: After briefly reviewing existing strategies for prior art search as well as machine learning methods for full text similarity search and its applications, we discuss our approach for computing the similarities between the patents using different feature extraction methods. These methods are then evaluated on an example dataset of patents including their citations, as well as a second dataset where relevant patents were identified by a patent attorney. Furthermore, based on this manually annotated dataset, we also assess the quality of the original citation process itself. A discussion of the relevance of the obtained results and a brief outlook conclude this manuscript.

\subsection{Related work}
Most research concerned with facilitating and improving the search for a patent's prior art has focused on automatically composing and extending the search queries. For example, a manually formulated query can be improved by automatically including synonyms for the keywords using a thesaurus \cite{magdy2011study,tannebaum2015patnet,Lupu2017book2,magdy2009exploring,wang2013semantic}. A potential drawback of such an approach, however, is that the thesaurus itself has to be manually curated and extended \cite{Zhang2016}. Another line of research focuses on pseudo-relevance feedback, where, given an initial search, the first $k$ search results are used to identify additional keywords that can be used to extend the original query \cite{mahdabi2012learning,ganguly2011patent,golestan2015term}. Similarly, past queries \cite{tannebaum2014using} or meta data such as citations can be used to augment the search query \cite{fujii2007enhancing,mahdabi2014effect,mahdabi2014query}. A recent study has also examined the possibility of using the \emph{word2vec} language model \cite{Mikolov2013a,Mikolov2013b,Mikolov2013c} to automatically identify relevant words in the search results that can be used to extend the query \cite{singh2016relevance}.

Approaches for automatically adapting and extending queries still require the patent examiner to manually formulate the initial search query. To make this step obsolete, heuristics can be used to automatically extract keywords from a given patent application \cite{mahdabi2011building,konishi2005query,verma2011applying} or a \emph{bag-of-words} (BOW) approach can be used to transform the entire text of a patent into a list of words that can then be used to search for its prior art \cite{verberne2009prior,bouadjenek2015study,xue2009transforming}. Often times, partial patent applications, such as an extended abstract, may already suffice to conduct the search \cite{bouadjenek2015study}. The search results can also be further refined with a graph-based ranking model \cite{mihalcea2004textrank} or by using the patents' categories to filter the results \cite{verma2011exploring}. Different prior art search approaches have previously been discussed and benchmarked within the CLEF-IP project, see e.g.\ \cite{piroi2010clef} and \cite{piroi2013overview}.

In our approach, detailed in the following sections, we also alleviate the required work and time needed to manually compose a search query by simply operating on the patent application's \emph{entire} text. However, instead of only searching the database for relevant keywords extracted from this text, we transform the texts of all other documents into numerical feature representations as well, which allow us to compute the full text similarities between the patent application and its possible prior art.

Calculating the similarity between texts is at the heart of a wide range of information retrieval tasks, such as search engine development, question answering, document clustering, or corpus visualization. Approaches for computing text similarities can be divided into similarity measures relying on word similarities and those based on document feature vectors \cite{gomaa2013survey}.

To compute the similarity between two texts using individual word similarities, the words in both texts first have to be aligned by creating word pairs based on semantic similarity and then these similarity scores are combined to yield a similarity measure for the whole text. Corley and Mihalcea \cite{Corley2005} propose a text similarity measure, where the most similar word pairs in two texts are determined based on semantic word similarity measures as implemented in the WordNet similarity package \cite{Patwardhan2003}. The similarity score of two texts is then computed as the weighted and normalized sum of the single word pairs' similarity scores. This approach can be further refined using greedy pairing \cite{lintean2012measuring}. Recently, instead of using WordNet relations to obtain word similarities, the similarity between semantically meaningful word embeddings, such as those created by the \emph{word2vec} language model \cite{Mikolov2013a}, was used. Kusner et al.~\cite{kusner2015word} defined the word mover's distance for computing the similarity between two sentences as the minimum distance the individual word embeddings have to move to match those of the other sentence. While similarity measures based on the semantic similarities of individual words are advantageous when comparing short texts, finding an optimal word pairing for longer texts is computationally very expensive and therefore these similarity measures are less practical in our setting, where the full texts of whole documents have to be compared.

To compute the similarity between longer documents, these can be transformed into numerical feature vectors, which serve as input to a similarity function. Rieck and Laskov \cite{Rieck2008} give a comprehensive overview of similarity measures for sequential data, some of which are widely used in information retrieval applications. Achananuparp et al.~\cite{Achananuparp2008} test some of these similarity measures for comparing sentences on three corpora, using accuracy, precision, recall, and rejection as metrics to evaluate how many of the retrieved documents are relevant in relation to the number of relevant documents missed. Huang \cite{Huang2008} use several of these similarity measures to perform text clustering on \textit{tf-idf} vectors.
Interested in how well similarity measures reproduce human similarity ratings, Lee et al.~\cite{lee2005empirical} create a text similarity corpus based on all possible pairs of 50 different documents rated by 83 students. They test different feature extraction methods in combination with four of the similarity measures described in Rieck and Laskov \cite{Rieck2008} and calculate the correlation of the human ratings with the resulting scoring. They conclude that using the cosine similarity, high precision can be achieved, while recall is still not satisfying.

Full text similarity measures have previously been used to improve search results for MEDLINE articles, where a two step approach using the cosine similarity measure between \emph{tf-idf} vectors in combination with a sentence alignment algorithm yielded superior results compared to the boolean search strategy used by PubMed \cite{lewis2006text}. The Science Concierge \cite{achakulvisut2016science} computes the similarities between papers' abstracts to provide content based recommendations, however it still requires an initial keyword search to retrieve articles of interest. The PubVis web application by Horn \cite{Horn2017}, developed for visually exploring scientific corpora, also provides recommendations for similar articles given a submitted abstract by measuring overlapping terms in the document feature vectors. While full text similarity search approaches have shown potential in domains such as scientific literature, only few studies have explored this approach for the much harder task of retrieving prior art for a new patent application \cite{moldovan2005latent}, where much less overlap between text documents is to be expected due to the usage of very abstract and general terms when describing new inventions. Specifically, document representations created using recently developed neural network language models such as \emph{word2vec} \cite{Mikolov2013a,Mikolov2013b,horn2017conecRepL4NLP} or \emph{doc2vec} \cite{Mikolov2014} were not yet evaluated on patent documents.

\section{Methods}
In order to study our hypothesis that the search for prior art can be improved by automatically determining, for a given patent application, the most similar documents contained in the database based on their full texts, we need to evaluate multiple approaches for comparing the patents' full texts and computing similarities between the documents. To do this, we test multiple approaches for creating numerical feature representations from the documents' raw texts, which can then be used as input to a similarity function to compute the documents' similarity.

All raw documents first have to be preprocessed by lower casing and removing non-alphanumeric characters.
The simplest way of transforming texts into numerical vectors is to create high dimensional but sparse \textit{bag-of-words} (BOW) vectors with \emph{tf-idf} features~\cite{Manning2008}. These BOW representations can also be reduced to their most expressive dimensions using dimensionality reduction methods such as \textit{latent semantic analysis} (LSA)~\cite{Landauer1998,moldovan2005latent} or \textit{kernel principal component analysis} (KPCA)~\cite{Schoelkopf1998,Mueller2001,scholkopf2002learning,Scholkopf2003}. Alternatively, the neural network language models (NNLM)~\cite{bengio2003neural} \textit{word2vec}~\cite{Mikolov2013a,Mikolov2013b} (combined with BOW vectors) or \textit{doc2vec}~\cite{Mikolov2014} can be used to transform the documents into feature vectors. All these feature representations are described in detail in the Supporting Information~\ref{S1a}.

Using any of these feature representations, the pairwise similarity between two documents' feature vectors $\mathbf{x}_i$ and $\mathbf{x}_j$ can be calculated using the cosine similarity:
\begin{align*}
\text{sim}\left(\mathbf{x}_i,\mathbf{x}_j\right) &= \frac{\mathbf{x}_{i}^\top\mathbf{x}_j}{\Vert \mathbf{x}_i \Vert \Vert \mathbf{x}_j\Vert},
\end{align*}
which is $1$ for documents that are (almost) identical, and $0$ (in the case of non-negative BOW feature vectors) or below $0$ for unrelated documents \cite{Crocetti2015, Huang2008, Yates1999}. Other possible similarity functions for comparing sequential data \cite{Rieck2008,Pele2011} are discussed in the Supporting Information~\ref{S1b}.

\section{Data}
Our experiments are conducted on two datasets, created using a multi-step process as briefly outlined here and further discussed in the Supporting Information~\ref{S2}. For ease of notation, we use the term patent when really referring to either a granted patent or a patent application.

We first obtained a patent corpus containing more than 100,000 patent documents from the \textit{Cooperative Patent Classification scheme} (\textsc{CPC}) category \textsc{A61} (\textit{medical or veterinary science and hygiene}), published between 2000 and 2015. From these documents, our first dataset was compiled, starting with the roughly 2,500 patents in the corpus published in 2015, which we will refer to as ``target patents'' in the remaining text. Each of the target patents cites on average 17.5 (standard deviation: $\pm$ 28.4) other patents in our corpus (i.e.\ published after 2000), which we also include in the dataset. Additionally, we randomly selected another 1,000 patents from the corpus, which were not cited by any of the selected target patents. This results in altogether 28,381 documents, which contain on average 13,530 ($\pm$ 18,750) words. From these documents, the first dataset was then created by pairing up the patents and assigning each patent pair a corresponding label: Each target patent is paired up with a) all the patents it cites, these patent pairs are assigned the label `cited', and b) the 1,000 patents not cited by any of the target patents, these patent pairs are labelled `random'. This first dataset consists of 2,470,736 patent pairs with a `cited/random' labelling.

The second dataset is created by obtaining additional, more consistent human labels from a patent attorney for a small subset of the first dataset. These labels should show which of the cited patents are truly relevant to the target patent and whether important prior art is missing from the search reports.
For ten of the target patents, we selected their respective cited patents as well as several random patents that either obtained a relatively high, medium, or low similarity score as computed with the cosine similarity on \emph{tf-idf} BOW features. These 450 patent pairs were then manually assigned `relevant/irrelevant' labels and constitute our second dataset.

\section{Evaluation}
A pair of patents should have a high similarity score if the two texts address a similar or almost identical subject matter, and a low score if they are unrelated. Furthermore, if two patent documents address a similar subject matter, then one document of said pair should have been cited in the search report of the other.
To evaluate the similarity computation with different feature representations, the task of finding similar patents can be modelled as a binary classification problem, where the samples correspond to pairs of patents. A patent pair is given a positive label, if one of the patents was cited by the other, and a negative label otherwise. We can then compute similarity scores for all pairs of patents and select a threshold for the score where we say all patent pairs with a similarity score higher than this threshold are relevant for each other while similarity scores below the threshold indicate the patents in this pair are unrelated. With a meaningful similarity measure, it should be possible to choose a threshold such that most patent pairs associated with a positive label have a similarity score above the threshold and the pairs with negative labels score below the threshold, i.e., the two similarity score distributions should be well separated. For a given threshold, we can compute the \textit{true positive rate} (TPR), also called \emph{recall}, and the \textit{false positive rate} (FPR) of the similarity measure. By plotting the TPR against the FPR for different decision thresholds, we obtain the graph of the \emph{receiver operating characteristic} (ROC) curve, where the \textit{area under the \textsc{ROC} curve} (\textsc{AUC}) conveniently translates the performance of the similarity measure into a number between $0.5$ (similarity scores assigned to patent pairs with a `cited' relationship and randomly paired patents are in the same range) and $1$ (semantically related patents receive consistently higher similarity scores than unrelated patent pairs). Further details on this performance measure can be found in the Supporting Information~\ref{S3}.

While the AUC is a very useful measure to select a similarity function based on which relevant and irrelevant patents can be reliably separated, the exact score also depends on characteristics of the dataset and may therefore seem overly optimistic \cite{saito2015precision}. Especially in our first dataset, many of the randomly selected patents contain little overlap with the target patents and can therefore be easily identified as irrelevant. With only a small fraction of the random pairs receiving a medium or high similarity score, this means that for most threshold values the FPR will be very low, resulting in larger AUC values. To give a further perspective on the performance of the compared similarity measures, we therefore additionally report the \emph{average precision} (AP) score for the final results.
For a specific threshold, precision is defined as the number of TP relative to the number of all returned documents, i.e., TP+FP. As we rank the patent pairs based on their similarity score, precision and recall can again be plotted against each other for $n$ different thresholds and the area under this curve can be computed as the weighted average of precision ($P$) and recall ($R$) for all $n$ threshold values \cite{zhu2004recall}:
\begin{align*}
AP = \sum_n (R_n - R_{n-1})P_n.
\end{align*}

\section{Results}
The aim of our study is to identify a robust approach for computing the full text similarity between two patents. To this end, in the following we evaluate different document feature representations and similarity functions by assessing how well the computed similarity scores are aligned with the labels of our two datasets, i.e.,\ whether a high similarity score is assigned to pairs that are labelled as \emph{cited} (\emph{relevant}) and low similarity scores to \emph{random} (\emph{irrelevant}) pairs. Furthermore, we examine the discrepancies between patents cited in a patent application's search report and truly relevant prior art. The data and code to replicate the experiments is available online.\footnote{\url{https://github.com/helmersl/patent_similarity_search}}

\subsection{Using full text similarity to identify cited patents}
The similarities between the patents in each pair contained in the cited/random dataset are computed using the different feature extraction methods together with the cosine similarity and the obtained similarity scores are then evaluated by computing the AUC with respect to the pairs' labels (Table~\ref{table:comp_sections_d2v}).
The similarity scores are computed using either the full texts of the patents to create the feature vectors, or only parts of the documents, such as the patents' abstracts or their claims, to identify which sections are most relevant for this task \cite{dhondt2010sections,bouadjenek2015study}.  Additionally, the results on this dataset using BOW feature vectors together with other similarity measures can be found in the Supporting Information~\ref{S4a}.
\begin{table}[!htb]
\centering
\caption{{\bf Evaluation results on the cited/random dataset.} AUC values when computing the cosine similarity with BOW, LSA, KPCA, \textit{word2vec}, and \textit{doc2vec} features constructed from different patent sections of the cited/random dataset.}
\begin{tabular}{lccc}
\toprule
\textbf{Features} & \multicolumn{3}{c}{\textbf{patent section: AUC}}\\ \midrule
  & \emph{full text} & \emph{abstract} & \emph{claims}\\ \cmidrule(l){2-4}
\emph{Bag-of-words} & \textbf{0.9560} & 0.8620 & 0.8656\\
\emph{LSA} & 0.9361 & 0.8579 & 0.8561\\
\emph{KPCA} & 0.9207 & 0.8377 & 0.8250\\
\emph{BOW + word2vec} & 0.9410 & 0.8618 & 0.8525\\
\emph{doc2vec} & 0.9314 & \textbf{0.8919} & \textbf{0.8898}\\
\bottomrule
\end{tabular}
\label{table:comp_sections_d2v}
\end{table}

The BOW features outperform the tested dimensionality reduction methods LSA and KPCA as well as the NNLM \textit{word2vec} and \textit{doc2vec} when comparing the patents' full texts (Table~\ref{table:comp_sections_d2v}). Yet, with AUC values greater than 0.9, all methods succeed in identifying cited patents by assigning the patents found in a target patent's search report a higher similarity score than those that they were paired up with randomly.
When only certain patent sections are taken into account, the NNLMs perform as good (\textit{word2vec}) or even better (\textit{doc2vec}) than the BOW vectors, and LSA performs well on the claims section as well. The comparably good performance, especially of \emph{doc2vec}, on individual sections is probably due to the fact that these feature representations are more meaningful when computed for shorter texts, whereas when combining the embedding vectors of too many individual words, the resulting document representation can be rather noisy.

\begin{figure*}[!htb]
\centering
\includegraphics[width=\linewidth]{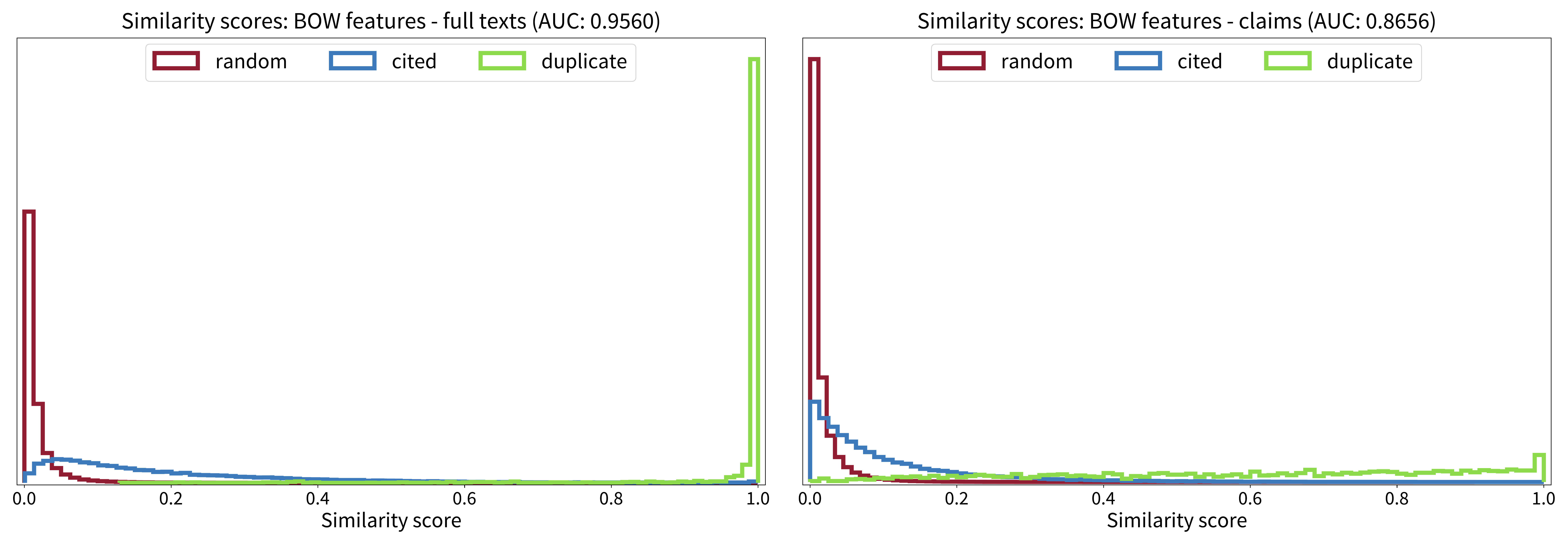}
\caption{{\bf Distributions of cosine similarity scores.} Similarity scores for the patent pairs are computed using BOW feature vectors generated either from full texts \emph{(left)} or only the claims sections \emph{(right)}.}\label{fig2}
\end{figure*}

When looking more closely at the score distributions obtained with BOW features on the patents' full texts as well as their claims sections (Fig~\ref{fig2}), it can be seen that when only using the claims sections, the scores of the duplicate patent pairs, instead of being clustered near $1$, range nearly uniformly between $0$ and $1$.
This can be explained by divisional applications and the fact that during the different stages of a submission process, most of the time only the claims section is revised (usually by weakening the claims), such that several versions of a patent application will essentially differ from each other only in their claims whereas abstract and description remain largely unchanged \cite{xue2009transforming,bouadjenek2015study}.

\subsection{Identifying truly relevant patents}
The search for prior art for a given patent application is in general conducted by a single person using mainly keyword searches, which might result in false positives as well as false negatives. Furthermore, as different patent applications are handled by different patent examiners, it is difficult to obtain a consistently labelled dataset. A more reliably labelled dataset would therefore be desirable to properly evaluate our automatic search approach. In the previous section, we showed that by computing the cosine similarity between feature vectors created from full patent texts we can identify patents that occur in the search report of a target patent. However, the question remains, whether these results translate to a real setting and if it is possible to find patents previously overlooked or prevent the citation of actually irrelevant patents.

To get an estimate of how many of the cited, as well as the patents identified through our automated approach, are truly relevant for a given target patent, we asked a patent attorney to label a small subsample of the first dataset. As the patent attorney labelled these patents very carefully, her decisions merit a high confidence and we therefore consider them as the ground truth when her ratings are in conflict with the citation labels.

Using this second, more reliably labelled dataset, we first assess the amount of (dis)agreement between the cited/random labelling, based on the search reports, and the relevant/irrelevant labelling, obtained from the patent attorney. We then evaluate the similarity scores computed for this second dataset to see whether our automated approach is indeed capable of identifying the truly relevant prior art for a new patent application.

\subsubsection*{Comparing the current citation process to the additional human labels}
To see if documents found in the search for prior art conducted by the patent office generally coincide with the documents considered relevant by our patent attorney, the confusion matrix as well as the correlation between the two human labellings is analysed. Please keep in mind that, in general, patent examiners can only assess the relevance of prior art that was actually found by the keyword driven search.

Taking the relevant/irrelevant labelling as the ground truth, the confusion matrix (Table~\ref{table:conf_matrix}) shows that 86~FP and 18~FN are produced by the patent examiner, which results in a recall of 0.78 and a precision score of 0.43. The large number of false positives can, in part, be explained by applicants being required by the USPTO to file so-called Information Disclosure Statements (IDS) including, according to the applicant, related background art \cite{uspto609}. The documents cited in an IDS are then included in the list of citations by the examiner, thus resulting in very long citations lists.
\begin{table}[!htb]
\centering
\caption{{\bf Confusion matrix for the dataset subsample.} The original cited/random labelling is compared to the more accurate relevant/irrelevant labels.}
\begin{tabular}{l c c}
\toprule
 & \textbf{cited}&\textbf{random} \\ \midrule
\textbf{relevant} & 65 &  18\\
\textbf{irrelevant} & 86 &  281\\
 \bottomrule
\end{tabular}
\label{table:conf_matrix}
\end{table}

To get a better understanding of the relationship between the cosine similarity computed using BOW feature vectors and the relevant/irrelevant as well as the cited/random labelling, we calculate their pairwise correlations using Spearman's $\rho$ (Table~\ref{table:simcoef_corr}).
The highest correlation score of 0.652 is reached between the relevant/irrelevant labelling and the cosine similarity, whereas Spearman's $\rho$ for the cosine similarity and the cited/random labels is much lower (0.501).
\begin{table}[!htb]
\centering
\caption{{\bf Correlations between labels and similarity scores on the dataset subsample.} Spearman's $\rho$ for the cosine similarity calculated with BOW feature vectors and the relevant/irrelevant and cited/random labelling.}
\begin{tabular}{l c c}
\toprule
 & \textbf{cited/random}&\textbf{relevant/irr.} \\ \midrule
\textbf{cosine (BOW)} &0.501 &  0.652 \\
\textbf{relevant/irr.} &0.592 &  ---\\
\bottomrule
\end{tabular}
\label{table:simcoef_corr}
\end{table}

\begin{figure}[!htb]
\centering
\includegraphics[width=\linewidth]{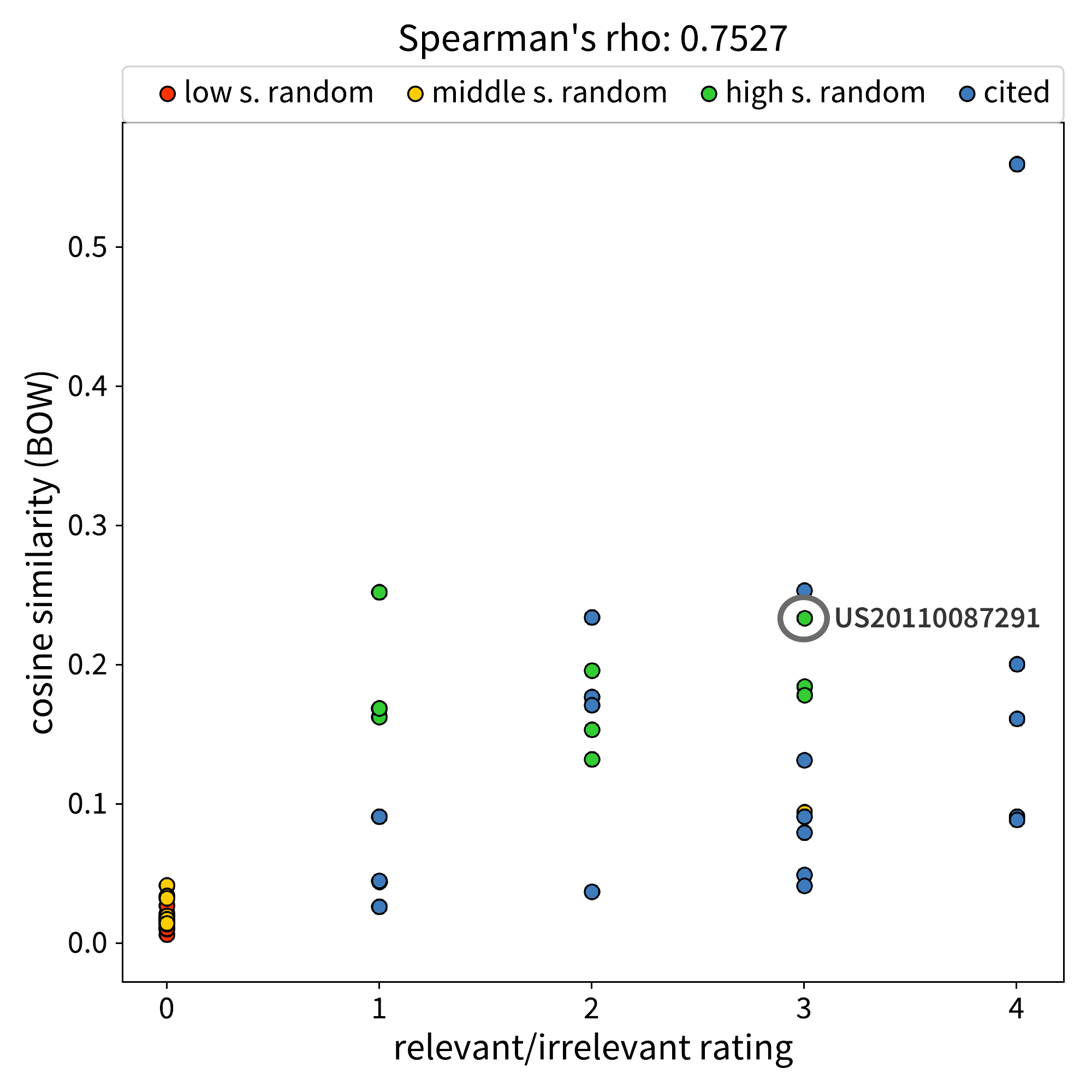}
\caption{{\bf Score correlation for the patent with ID US20150018885.} A false negative (ID~US20110087291) caught by the cosine similarity is circled in gray.}\label{fig3}
\end{figure}

When plotting the cosine similarity and the relevant/irrelevant labelling against each other for individual patents (e.g.\ Fig~\ref{fig3}), in most cases, the scorings agree on whether a patent is relevant or not for the target patent. Yet it is worthwhile to inspect some of the outliers to get a better understanding of the process. In the Supporting Information~\ref{S4b} we discuss two false positives, one produced by our approach and one found in a patent's search report. More problematic, however, are false negatives, i.e.,~prior art that was missed when filing the application. For the target patent with ID~US20150018885 our automated approach would have discovered a relevant patent, which was missed by the search performed by the patent examiner (Fig~\ref{fig3}).
The patent with ID~US20110087291 must be considered as relevant for the target patent, because both describe rigid bars that are aimed at connecting vertebrae for stabilization purposes with two anchors that are screwed into the bones. While in the target patent, the term \textit{bone anchoring member} is used, the same part of the device in patent US20110087291 is called \textit{connecting member}, which is a more abstract term. Moreover, instead of talking about a \textit{connecting bar}, as it is done in the target patent, the term \textit{elongate fusion member} is used in the other patent application.

\subsubsection*{Using full text similarity to identify relevant patents}
In order to systematically assess how close the similarity score ranking can get to the one of the patent attorney (relevant/irrelevant) compared to the one of the patent office examiners (cited/random), the experiments performed on the first dataset with respect to the cited/random labelling were again conducted on this dataset subsample. For the analysis, it is important to bear in mind that this dataset is different from the one used in the previous experiments, as it only consists of the 450 patent pairs scored by the patent attorney. For each of the feature extraction methods, it was assessed how well the cosine similarity could distinguish between the relevant and irrelevant as well as the cited and random patent pairs of this smaller dataset.

The AUC and AP values achieved with the different feature representations on both labellings as well as, for comparison, on the original dataset, are reported in Table~\ref{table:auc_results_summary}.
\begin{table*}[!hbt]
\centering
\caption{{\bf Summary of evaluation results.} AUC and average precision (AP) scores for the different feature extraction methods on the dataset subsample with cited/random and relevant/irrelevant labelling, as well as the full dataset.}
\begin{tabular}{l c c c c c c}
\toprule
 \textbf{Features} &\multicolumn{3}{c}{\textbf{AUC}}&\multicolumn{3}{c}{\textbf{AP}}\\ \midrule
 &\multicolumn{2}{c}{\emph{ subsample}} &\emph{full }&\multicolumn{2}{c}{\emph{ subsample}} &\emph{full }\\ \cmidrule(l){2-4}\cmidrule(l){5-7}
                    & {relevant}    & {cited}   &  {cited}          & {relevant} & {cited} & {cited}\\ \cmidrule(l){2-3}\cmidrule(l){4-4}\cmidrule(l){5-6}\cmidrule(l){7-7}
\emph{Bag-of-words} & 0.8118        &  0.8063   & \textbf{0.9560}   & 0.5274 & 0.7095 & \textbf{0.4705}\\
\emph{LSA}          & 0.7798        &  0.7075   & 0.9361            & 0.4787 & 0.5921 & 0.3257\\
\emph{KPCA}         & 0.7441        &  0.6740   & 0.9207            & 0.4721 & 0.5832 & 0.2996\\
\emph{BOW + word2vec} & \textbf{0.8408} & \textbf{0.8544} & 0.9410  & \textbf{0.5443} & \textbf{0.7354} & 0.4019\\
\emph{doc2vec}      & 0.7658        &  0.8138   & 0.9314            & 0.4749 & 0.6829 & 0.3121\\
\bottomrule
\end{tabular}
\label{table:auc_results_summary}
\end{table*}
On this dataset subsample, the AUC w.r.t.~the cited/random labelling is much lower than in the previous experiment on the larger dataset (0.806 compared to 0.956 for BOW features), which can be in part explained by the varying number of easily identifiable negative samples and their impact on the FPR: The full cited/random dataset contains many more low-scored random patents than the relevant/irrelevant subsample, where we included an equal amount of low- and high-scored random patents for each of the ten target patents.
Yet, for most feature representations, the performance is better for the relevant/irrelevant than for the cited/random labelling of the dataset subsample, and the best results on the relevant/irrelevant labelling are achieved using the combination of BOW vectors and \emph{word2vec} embeddings as feature vectors.

\section{Discussion}
The search for prior art for a given patent application is currently based on a manually conducted keyword search, which is not only time consuming but also prone to mistakes yielding both false positives and, more problematically, false negatives. In this paper, an approach for automating the search for prior art was developed, where a patent application's \emph{full} text is automatically compared to the patents contained in a database, yielding a similarity score based on which the patents can be ranked from most similar to least similar. The patents whose similarity scores exceed a certain threshold can then be suggested as prior art.

Several feature extraction methods for transforming documents into numerical vectors were evaluated on a dataset consisting of several thousand patent documents.
In a first step, the evaluation was performed with respect to the distinction between cited and random patents, where cited patents are those included in the given target patent's search report and random patents are randomly selected patent documents that were not cited by any of the target patents. We showed that by computing the cosine similarity between feature vectors created from full patent texts, we can reliably identify patents that occur in the search report of a target patent. The best distinction between these cited and random patents on the full corpus could be achieved when computing the cosine similarity using the well-established \textit{tf-idf} BOW features, which is conceptually the method most closely related to a regular keyword search.

To examine the discrepancies between the computed similarity scores and cited/random labels, we obtained additional and more reliable labels from a patent attorney to identify truly relevant patents. As illustrated by Tables~\ref{table:simcoef_corr} and \ref{table:auc_results_summary}, the automatically calculated similarities between patents are closer to the patent attorney's relevancy scoring than to the cited/random labellings obtained from the search report.
The comparison of different feature representations on the smaller dataset not only showed that the same feature extraction method reaches different AUCs for the two labellings, but also that the feature extraction method that best distinguishes between cited and random patents on the full corpus (BOW) was outperformed on the relevant/irrelevant dataset by the combination of \textit{tf-idf} BOW feature vectors with \textit{word2vec} embeddings. This again indicates that the keyword search is missing patents that use synonyms or more general and abstract terms, which can be identified using the semantically meaningful representations learned by a NNLM. Therefore, with our automated similarity search, we are able to identify the truly relevant documents for a given patent application.

Most importantly, we gave an example where the cosine similarity caught a relevant patent originally missed by the patent examiner (Fig~\ref{fig3}). As discussed at the beginning of this paper, missing a relevant prior art document in the search is a serious issue, as this might lead to an erroneous grant of a patent with profound legal and financial implications for both the applicant as well as competitors.

Consequently, our findings show that the search for prior art for a given patent application, and thereby the citation process, can be greatly enhanced by a precursory similarity scoring of the patents based on their full texts. With our NLP based approach we would not only greatly accelerate the search process, but, as shown in our empirical analysis, our method could also improve the \emph{quality} of the results by reducing the number of omitted yet relevant documents.

Given the so far unsatisfying precision (0.43) and recall (0.78) values of the standard citation process compared to the relevancy labellings provided by our patent attorney, in the future it is clearly desirable to focus on improving the separation of relevant and irrelevant instead of cited and random patents. Our results on the small relevant/irrelevant dataset, while very encouraging, should only be considered as a first indicative step; clearly the creation of a larger dataset, reliably labelled by several experts, will be an essential next step for any further evaluation.

While we have demonstrated that our search approach is capable of identifying FP and FN w.r.t.\ the documents cited in a patent's original search report, it is not clear whether this original search for prior art was always conducted using any of the more sophisticated IR approaches discussed in the related works section at the beginning of the paper, i.e., going beyond a basic manual keyword search. Therefore, a future step in the evaluation of our search approach would be to benchmark our methods against these existing IR techniques specifically developed for the prior art search, for example, using the CLEF-IP datasets \cite{piroi2010clef,piroi2013overview}.

Furthermore, the methods discussed within this paper should also be applied to documents from other CPC classes to assess the quality of the automatically generated search results in domains other than medical or veterinary science and hygiene. Additionally considering the (sub)categories of the patents as features when conducting the search for prior art also seems like a promising step to further enhance the search results \cite{verma2011exploring,magali2016review}.

It should also be evaluated how well these results translate to patents filed in other countries \cite{Piroi2017book4,Lupu2017book3}, especially if these patents were automatically translated using machine translation methods \cite{Tinsley2017book16,Diallo2017bookfuture}. Here it may also be important to take a closer look at similarity search results obtained by using only the texts from single patent sections. As related work has shown \cite{bouadjenek2015study,dhondt2010sections}, an extended abstract and description may often suffice to find prior art. This can speed up the patent filing process, as all relevant prior art can already be identified early in the patent application process, thereby reducing the number of duplicate submissions with only revised (i.e.\ weakened) claims. However, as patents filed in different countries have different structures, these results might not directly translate to, e.g., patents filed with the European Patent Office.

It might also be of interest to compare other NNLM based feature representations for this task, e.g.,~by combining the \emph{word2vec} embeddings with a convolutional neural network \cite{arras2016explaining,arras2017relevant}. To better adapt a similarity search approach to patents from other domains, it could also be advantageous to additionally take into account image based similarities computed from the sketches supplied in the patent documents \cite{Alberts2017book1,lupu2013patent}.

An important challenge to solve furthermore is how an exhaustive comparison of a given patent application to all the millions of documents contained in a real world patent database could be performed efficiently. Promising approaches for speeding up the similarity search for all pairs in a set~\cite{Bayardo2007} should be explored for this task in future work.

The search for a patent's prior art is a particularly difficult problem, as patent applications are purposefully written in a way that is to create little overlap with other patents, as only by distinguishing the invention from others, a patent application has a chance of being granted \cite{Andersson2017book9}. By showing that our automated full text similarity search approach successfully improves the search for a patent's prior art, consequently these methods are also promising candidates for enhancing other document searches, such as identifying relevant scientific literature.

\subsection*{Acknowledgements}
{\small This work was supported by the Federal Ministry of Education and Research (BMBF) for the Berlin Big Data Center BBDC (01IS14013A) and Berlin Center for Machine Learning BZML (01IS18037I), as well as the Institute of Information \& Communications Technology Planning \& Evaluation (IITP) grant funded by the Korea government (No. 2017-0-00451). Pfenning, Meinig \& Partner mbB provided support in the form of salaries for authors FB and TO. The funders had no role in study design, data collection and analysis, decision to publish, or preparation of the manuscript.}

\subsection*{Author contributions statement}
{\small FH, FB, and KRM discussed and conceived the experiments, LH conducted the experiments, FB and TO labelled the subsample of the dataset. All authors wrote and reviewed the manuscript. Correspondence to LH, FH, and KRM.}

\bibliography{patents}
\bibliographystyle{plainnat}

\appendix
\clearpage
\onecolumn
\section{Supporting Information: Methods}

\subsection{Feature representations of text documents}\label{S1a}

\subsubsection*{Tf-Idf BOW features}
Given $D$ documents with a vocabulary of size $L$, each text is transformed into a \emph{bag-of-words} (BOW) feature vector $\mathbf{x}_k \in \mathbb{R}^L \;\forall k \in 1...D$ by first computing a normalized count, the \emph{term frequency} (tf), for each word in a text, and then weighting this by the word's \emph{inverse document frequency} (idf) to reduce the influence of very frequent but inexpressive words that occur in almost all documents (such as `and' and `the')~\cite{Manning2008}. The idf of a term $w$ is calculated as the logarithm of the total number of documents, $|D|$, divided by the number of documents that contain term $w$, i.e.
\begin{align*}
 \text{idf}\,(w) &= \log {|D|\over |\{d \in D\text{ : }w \in d\}|}.
 \end{align*}
The entry corresponding to the word $w$ in the feature vector $\mathbf{x}_k$ of a document $k$ is then
\begin{align*}
\mathbf{x}_{k}(w) &= \text{tf}_k(w) \cdot \text{idf}\,(w).
\end{align*}
Instead of using the term frequency, a binary entry in the feature vector for each word occurring in the text might often suffice. Furthermore, the final \emph{tf-idf} vectors can be normalized by dividing them e.g.~by the maximum or the length of the respective vector:
\begin{align*}
\mathbf{\tilde x}_{k} = {\mathbf{x}_{k} \over \max_w \mathbf{x}_{k}(w)}\;\; &\text{  or  } \;\; \mathbf{\tilde x}_{k} = {\mathbf{x}_{k} \over \Vert \mathbf{x}_k \Vert}.
\end{align*}

\subsubsection*{LSA and KPCA}
Transforming the documents in the corpus into BOW vectors leads to a high-dimensional but sparse feature matrix. These feature representations can be reduced to their most expressive dimensions, which helps to reduce noise in the data and create more overlap between vectors. For this, we experiment with both \textit{latent semantic analysis} (LSA)~\cite{Landauer1998} and \textit{kernel principal component analysis} (KPCA)~\cite{Schoelkopf1998}.

\textsc{LSA} represents a word's meaning as the average of all the passages the word appears in, and a passage, such as a document, as the average of all the words it contains. Mathematically, a singular value decomposition (\textsc{SVD}) of the BOW feature matrix $X \in \mathbb{R}^{D\times L}$ for the respective corpus is performed. The original data points can then be projected onto the vectors corresponding to the $l$ largest singular values of matrix $X$, yielding a lower-dimensional representation $\hat{X} \in \mathbb{R}^{D \times l}$, where $l < L$. Choosing a dimensionality $l$ that is smaller than the original dimension $L$ is assumed to lead to a deeper abstraction of words and word sequences and to give a better approximation of their meaning~\cite{Landauer1998}.

Similarly, \textsc{KPCA}~\cite{Schoelkopf1998, Mueller2001} performs an SVD of a linear or non-linear kernel matrix $K \in \mathbb{R}^{D \times D}$ to obtain a low dimensional representation of the data, again based on the eigenvectors corresponding to the largest eigenvalues of this matrix. While we have studied different Gaussian kernels, we found that good results could already be obtained using the linear kernel $K = XX^\top$.

When reducing the dimensionality of the BOW feature vectors with LSA and KPCA, four embedding dimensions (100, 250, 500 and 1000) were tested and the best performance on the full texts was achieved using 1000 dimensions. As the dataset subsample contains only 450 patent pairs, here the best results with LSA and KPCA were achieved using only 100 dimensions.

\subsubsection*{Combining BOW features with word2vec embeddings}
One shortcoming of the BOW vectors is that semantic relationships between words, such as synonymy, as well as word order, are not taken into account. This is due to the fact that each word is associated with a single dimension in the feature vector and therefore the distances between all words are equal. The aspect of synonymy is especially relevant for patent texts, where very abstract and general terms are used for describing an invention in order to assure a maximum degree of coverage. For instance, a term like \textit{fastener} might be preferred over the usage of the term \textit{screw}, as it includes a wider range of material and therefore gives a better protection against infringement. Thus, patent texts tend to contain neologisms and abstract words that might even be unique in the corpus. To account for this variety in a keyword search is especially tedious and prone to errors as the examiner has to search for synonyms at different levels of abstraction or rely on a thesaurus, which would then need to be kept up-to-date \cite{Zhang2016}. Even the BOW approach could in this case only capture the similarity between the patent texts if there is overlap between the words in the context around a synonym. An approach specifically developed to overcome these restrictions are \textit{neural network language models} (\textsc{NNLM})~\cite{bengio2003neural}, which aim at representing words or documents by semantically meaningful vectorial embeddings.

A NNLM that recently received a lot of attention is \textit{word2vec}. Its purpose is to embed words in a vector space based on their contexts, such that terms appearing in similar contexts are close to each other in the embedding space w.r.t.~the cosine similarity~\cite{Mikolov2013a, Mikolov2013b,horn2017conecRepL4NLP}. Given a text corpus, the word representations are obtained by training a neural network that learns from the local contexts of the input words in the corpus. The embedding is then given by the learned weight matrix. Mikolov et al.~\cite{Mikolov2013a} describe two different network architectures for training the \textit{word2vec} model, namely the \textit{continuous bag-of-words} (CBOW) and the \textit{skip-gram} model. The first one learns word representations by predicting a target word based on its context words and the latter one by predicting the context words for the current input word. As the skip-gram model showed better performance in analogy tasks~\cite{Mikolov2013a, Mikolov2013b, Mikolov2013c} it is used in this paper.\footnote{Analogy tasks aim at finding relations such as \textit{\textbf{A} is to \textbf{B} as \textbf{C} is to $\_\_$}. For instance, in the relation \textit{\textbf{good} is to \textbf{better} as \textbf{bad} is to $\_\_$}, the correct answer would be \textit{\textbf{worse}}.}

To make use of the information learned by the \textit{word2vec} model for each word in the corpus vocabulary $L$, the trained word embeddings have to be combined to create a document vector for each patent text. To this end, the dot product of each document's BOW vector with the word embedding matrix $W \in \mathbb{R}^{L \times r}$, containing one $r$-dimensional word embedding per row, is calculated. For each document represented by a BOW vector $\mathbf{x}_k \in \mathbb{R}^{L}$, this results in a new document vector $\tilde{\mathbf{x}}_k \in \mathbb{R}^{r}$, which corresponds to the sum of the \emph{word2vec} embeddings of the terms occurring in the document, weighted by their respective \textit{tf-idf} scores. Combining the BOW vectors and the word embeddings thus comes along with a dimensionality reduction of the document vectors, while their sparseness is lost.

For the \textit{word2vec} model we use a standard setting from the literature (i.e.~the embedding dimension $r$ was set to $200$, the window size $c$ as well as the minimum frequency to 5 and negative sampling was performed using 13 noise words) \cite{Mikolov2013a, Mikolov2013b}.

\subsubsection*{Doc2vec representations}
With \textit{doc2vec}, Le and Mikolov \cite{Mikolov2014} extend the \textit{word2vec} model to directly represent word sequences of arbitrary lengths, such as sentences, paragraphs or even whole documents, by vectors. To learn the representations, word and paragraph vectors are trained simultaneously for predicting the next word for different contexts of fixed size sampled from the paragraph such that, at least in small contexts, word order is taken into account. Words are mapped to a unique embedding in a matrix~$W \in \mathbb{R}^{L\times r}$ and paragraphs to a unique embedding in a matrix~$P  \in \mathbb{R}^{D\times r}$. In each training step, paragraph and word embeddings are combined by concatenation to predict the next word given a context sampled from the respective paragraph.
After training, the \textit{doc2vec} model can be used to infer the embedding for an unseen document by performing gradient descent on the document matrix $P$ after having added more rows to it and holding the learned word embeddings and softmax weights fixed~\cite{Mikolov2014}.

For the \emph{doc2vec} model, we explored the parameter values 50, 100, 200 and 500 for the embedding dimension $r$ of the document vectors on the cited/random dataset in preliminary experiments, with the best results achieved with $r = 50$. The window size was set to 8, the minimum word count to 5, and the model was trained for 18 iterations. When training the model, the target patents were excluded from the corpus to avoid overfitting. Their document vectors were then inferred by the model given the learned parameters before computing the similarities to the other patents.

\subsection{Functions for measuring similarity between text documents}\label{S1b}
Transforming the patent documents into numeric feature vectors allows to assess their similarity with the help of mathematical functions. Rieck and Laskov \cite{Rieck2008} give a comprehensive overview on vectorial similarity measures for the pairwise comparison of sequential data. These can be divided into three main categories, namely \textit{kernels}, \textit{distance functions}, and \textit{similarity coefficients}. Their formulas are shown in Table~\ref{table:simmeasures} and the notation is consistent with the one in the paper. Here, $w$ corresponds to a word in the vocabulary~$L$ of the corpus, and $\Phi_w\left(x\right)$ maps each word $w \in L$ to its normalized and weighted count in sequence $x$, i.e.~to its \textit{tf-idf} value. The similarity functions will be briefly described in the following, while further details can be found in the original publication~\cite{Rieck2008}.
\begin{table}[!ht]
\centering
\caption{\textbf{Overview of similarity measures for sequential data~\cite{Rieck2008}.}}
\begin{tabular}{lc}
\toprule
\multicolumn{2}{c}{\textbf{Similarity coefficients}}\\
\midrule
Cosine & ${\sum_{w \in L} \Phi_w \left(x_i\right)\Phi_w\left(x_j\right) \over \sqrt{\sum_w \Phi_w \left(x_i\right)^2}\sqrt{\sum_w \Phi_w \left(x_j\right)^2}}$ \\[2ex]
Braun-Blanquet & ${a \over \max\left(a + b, a + c\right)}$ \\[2ex]
Czekanowski, S{\o}rensen-Dice & ${2a \over \left(2a + b + c\right)}$ \\[2ex]
Jaccard & ${a \over \left(a + b + c\right)}$ \\[2ex]
Kulczynski & ${a \over 2\left(a + b\right)} + {a \over 2\left(a + c\right)}$ \\[2ex]
Otsuka, Ochiai & ${a \over \sqrt{\left(a+b\right)\left(a+c\right)}}$ \\[2ex]
Simpson & ${a \over \min\left(a + b, a + c\right)}$ \\[2ex]
Sokal-Sneath, Anderberg & ${a \over \left(a + 2\left(b + c\right)\right)}$\\[2ex]
\midrule
 \multicolumn{2}{c}{\textbf{Kernel functions}} \\
 \midrule
 Linear &  $\sum_{w \in L} \Phi_w \left(x_i\right)\Phi_w\left(x_j\right)$\\[2ex]
  Gaussian  & $\exp\left({-d\left(x_i, x_j\right)^2 \over 2\sigma^2}\right)$\\[2ex]
Histogram intersection  & $\sum_{w \in L} \min\left(\Phi_w\left(x_i\right),\Phi_w\left(x_j\right)\right)$\\[2ex]
Polynomial &  $\left(\sum_{w \in L} \Phi_w \left(x_i\right)\Phi_w\left(x_j\right) + \Theta\right)^p$\\[2ex]
Sigmoidal & $\tanh\left(\sum_{w \in L} \Phi_w \left(x_i\right)\Phi_w\left(x_j\right) + \Theta\right)$\\[2ex]
\midrule
 \multicolumn{2}{c}{\textbf{Distance functions}} \\
\midrule
Canberra & $\sum_{w \in L}{\vert\Phi_w \left(x_i\right)-\Phi_w\left(x_j\right)\vert \over \Phi_w \left(x_i\right)+\Phi_w\left(x_j\right)}$ \\[2ex]
Chebyshev & $\max_{w \in L}\vert \Phi_w \left(x_i\right)-\Phi_w\left(x_j\right) \vert$\\[2ex]
Euclidean &  $\sum_{w \in L} \vert \Phi_w \left(x_i\right)-\Phi_w\left(x_j\right) \vert^2$\\ [2ex]
Geodesic & $\arccos \sum_{w \in L} \Phi_w \left(x_i\right)\Phi_w\left(x_j\right)$\\[2ex]
Hellinger$^2$ & $\sum_{w \in L} \left(\sqrt{\Phi_w\left(x_i\right)}  - \sqrt{\Phi_w\left(x_j\right)}\right)^2$ \\[2ex]
Jensen-Shannon & $\sum_{w \in L} H\left(\Phi_w\left(x_i\right),\Phi_w\left(x_j\right)\right)$ \\[2ex]
Manhattan & $\sum_{w \in L} \vert \Phi_w \left(x_i\right)-\Phi_w\left(x_j\right) \vert$ \\ [2ex]
Minkowski$^p$ & $\sum_{w \in L} \vert \Phi_w \left(x_i\right)-\Phi_w\left(x_j\right) \vert^p$\\ [2ex]
$\chi^2$ & $\sum_{w \in L}{\left(\Phi_w \left(x_i\right)-\Phi_w\left(x_j\right)\right)^2 \over \Phi_w \left(x_i\right)+\Phi_w\left(x_j\right) }$ \\  [2ex]
  \bottomrule
\end{tabular}
\label{table:simmeasures}
\end{table}
The general idea for the comparison of two sequences is that the more overlap they show with respect to their subsequences, the more similar they are. When transforming texts into BOW features, a subsequence corresponds to a single word. Two sequences~$x_i$ and~$x_j$ can thus be compared based on the normalized and weighted counts of the subsequences stored in the respective feature vectors $\mathbf{x}_i$ and $\mathbf{x}_j$.

\paragraph{Kernel functions} The first group of similarity measures Rieck and Laskov \cite{Rieck2008} discuss are kernel functions. They implicitly map the feature vectors into a possibly high or even infinite dimensional feature space, where the kernel can be expressed as a dot product. A kernel $k$ thus has the general form
\begin{equation}
k\left( \mathbf{x}_i, \mathbf{x}_j \right) = \langle f(\mathbf{x}_i), f(\mathbf{x}_j) \rangle, \nonumber
\end{equation}
where $f$ maps the vectors into the kernel feature space. The advantage of the kernel function is that it avoids the explicit calculation of the vectors' high dimensional mapping and allows to obtain the result in terms of the vectors' representation in the input space instead~\cite{Scholkopf2003,scholkopf2002learning}.

\paragraph{Distance functions} The distance functions described in Rieck and Laskov \cite{Rieck2008} are so-called \textit{bin-to-bin} distances~\cite{Pele2011}. This means that they compare each component of the vector to its corresponding component in the other one, e.g.~by subtracting the respective word counts and summing the subtractions for all words in the vocabulary. Unlike similarity measures, the distance measures are higher the more different the compared sequences are but can be easily transformed into a similarity measure by multiplying the result with $-1$, for example.

\paragraph{Similarity coefficients} Similarity coefficients were designed for the comparison of binary vectors and, instead of expressing metric properties, they assess similarity by comparing the number of matching components between two sequences. More precisely, for calculating the similarity of two sequences $x_i$ and $x_j$, they use three variables \textit{a}, \textit{b} and \textit{c}, where \textit{a} corresponds to the number of components contained in both $x_i$ and $x_j$, \textit{b} to the number of components contained in $x_i$ but not in $x_j$, and \textit{c} to the number of components contained in $x_j$ but not in $x_i$. In the case of BOW vectors, which are not inherently binary, the three variables can be expressed as follows:
\begin{eqnarray}
 a &=& \sum_{w\in L} \min \left(\Phi_w(x_i), \Phi_w(x_j)\right), \nonumber \\
 b &=& \sum_{w\in L} \left(\Phi_w(x_i) - \min \left(\Phi_w(x_i), \Phi_w(x_j)\right) \right), \nonumber\\
 c &=& \sum_{w\in L} \left(\Phi_w(x_j) - \min \left(\Phi_w(x_i), \Phi_w(x_j)\right) \right). \nonumber
\end{eqnarray}

\section{Supporting Information: Data} \label{S2}

\subsection*{Patent corpus}
To evaluate the different methods for computing document similarities on real world data, an initial patent corpus was obtained from a patent database.
This corpus consists of over 100,000 patent grants and applications published at the \textit{United States Patent and Trademark Office} (\textsc{USTPO}) between 2000 and 2015.

\begin{figure}[!ht]
\centering
\includegraphics[width=0.65\textwidth]{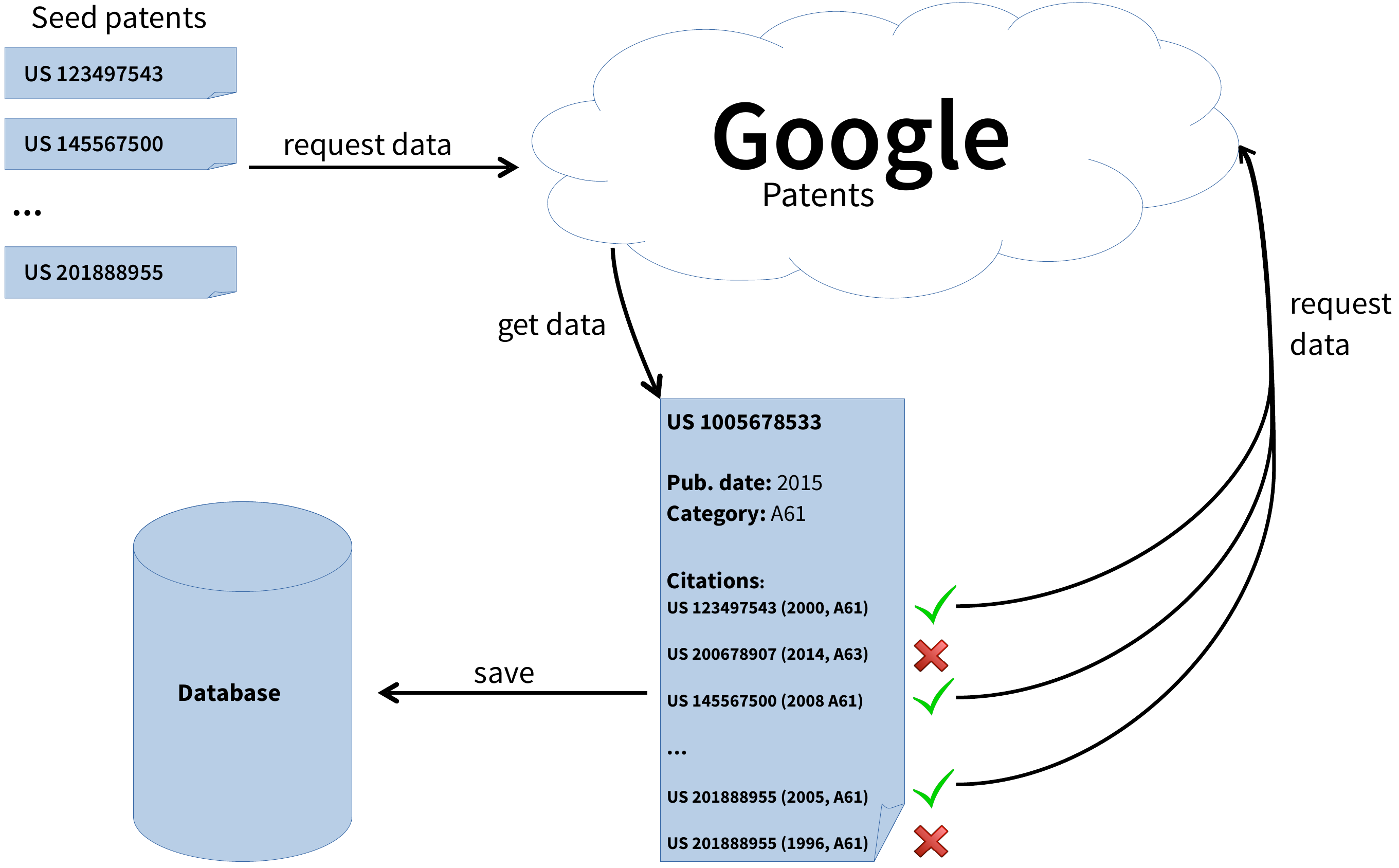}
\caption{\textbf{Illustration of the crawling process.}}\label{fig5}
\end{figure}

We create such a patent corpus (by crawling \textsc{Google Patents}\footnote{\url{https://www.google.de/patents}}) as illustrated in Fig~\ref{fig5}.
To get a more homogeneous dataset, only patents of the category \textsc{A61} (\textit{medical or veterinary science and hygiene}) according to the \textit{Cooperative Patent Classification scheme} (\textsc{CPC}) were included in our corpus. Another important criterion for including a patent document in our initial patent corpus was that its search report, i.e.~the prior art cited by the examiner, had to be available from the database. Starting with 20 manually selected (randomly chosen) seed patents published in 2015, the patent corpus was iteratively extended by including the seed patents' citations if they were published after 1999 and belonged to the category \textsc{A61}. The citations of these patents were then again checked for publication year and category and included if they fulfilled the respective conditions.

\subsubsection*{Structure of the crawled dataset}
\begin{figure}[!ht]
\centering
\includegraphics[width=0.57\textwidth]{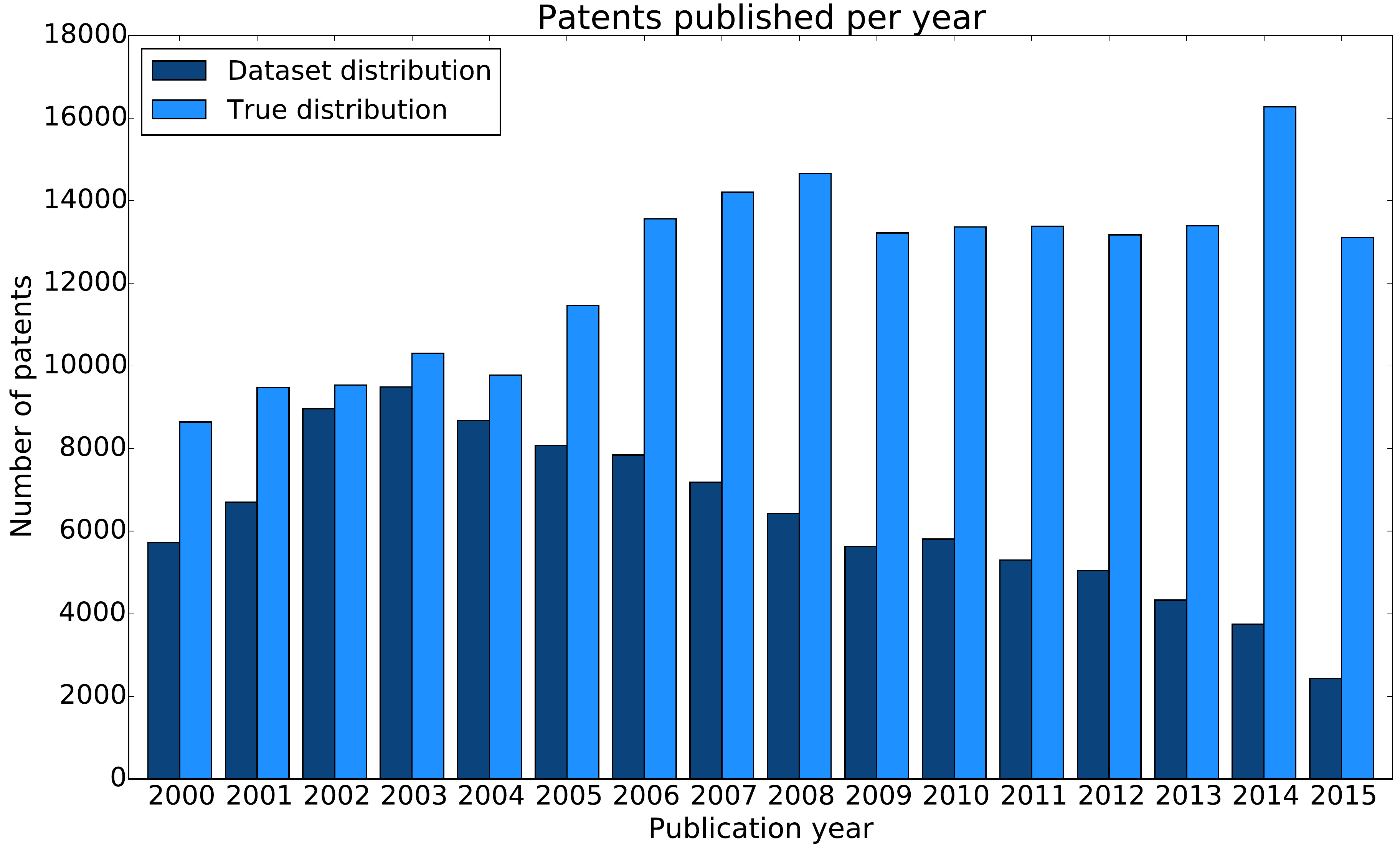}
\caption{\textbf{Number of published patents per year in the crawled dataset compared to the true distribution of patents published per year.}}\label{fig6}
\end{figure}

Comparing the distribution of patents published per year in the dataset and the total amount of patents filed between 2000 and 2015 at the \textsc{USTPO} (Fig~\ref{fig6}), it can be seen that the distribution in the dataset is not representative. The peak in 2003 and the fact that there are less and less patents with a publication date in the following years is most probably a result of the crawling strategy. Given that we started with some patents filed in 2015 and then subsequently crawled the citations, published in the past, explains the low amount of patents published in more recent years in the dataset.

\begin{figure}[!ht]
\centering
\includegraphics[width=0.57\textwidth]{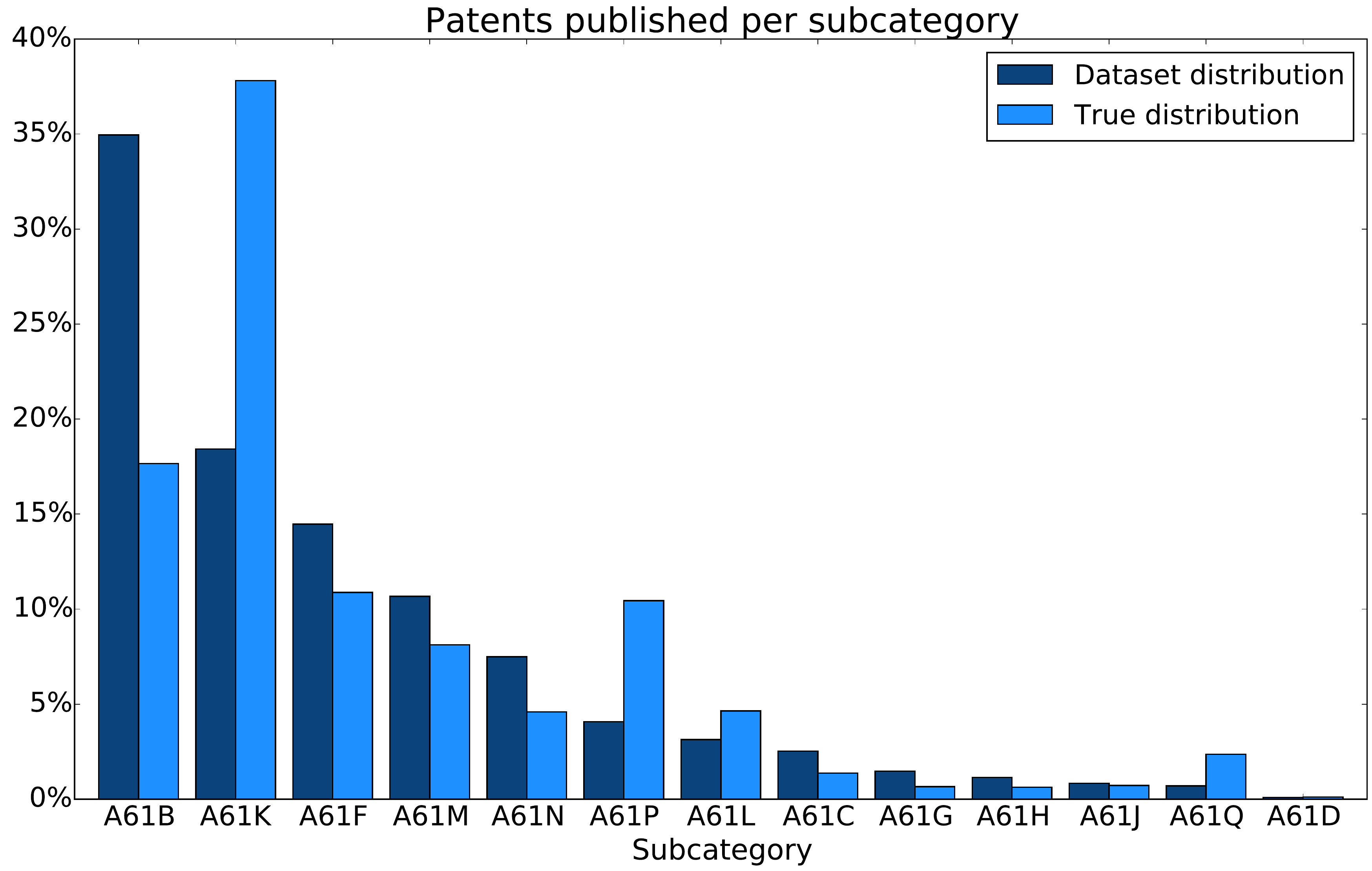}
\caption{\textbf{Subcategory distribution for crawled and published patents from 2000 to 2015.} Subcategory names can be found in Table~\ref{table:cpc_table}.}\label{fig7}
\end{figure}

The same holds for the subcategory distribution displayed in Fig~\ref{fig7}. While the most prominent subcategory in our dataset is \textsc{A61B}, the most frequent subcategory is actually \textsc{A61K}. The bias for subcategory \textsc{A61B} is due to the fact that several seed patents belonged to it.

\begin{table}[!ht]
\centering
\caption{ \textbf{CPC table for the subcategories of class \textbf{A61} (\textsc{medical or veterinary science and hygiene}).}}
\begin{footnotesize}
\begin{tabular}{l  l}
\toprule
\multirow{3}{*}{\textbf{A61B}}
& Diagnosis \\
& Surgery \\
& Identification \\ \midrule
\multirow{2}{*}{\textbf{A61C}}
& Dentistry \\
& Apparatus or methods for oral or dental hygiene \\ \midrule
\textbf{A61D}
& Veterinary instruments, implements, tools, or methods\\ \midrule
\multirow{8}{*}{\textbf{A61F}}
& Filters implantable into blood vessels \\
& Prostheses \\
& Devices providing patency to or preventing collapsing of tubular structures of the body, e.g. stents\\
& Orthopaedic, nursing or contraceptive devices \\
& Fomentation\\
& Treatment or protection of eyes or ears\\
& Bandages, dressings or absorbent pads \\
& First-aid kits  \\ \midrule
\multirow{4}{*}{\textbf{A61G}}
& Transport or accomodation for patients \\
& Operating tables or chairs \\
& Chairs for dentistry \\
& Funeral devices\\ \midrule
\multirow{4}{*}{\textbf{A61H}}
& Physical therapy apparatus, e.g. devices for locating or stimulating reflex points in the body \\
& Artificial respiration \\
& Massage\\
& Bathing devices for special therapeutic or hygienic purposes or specific parts of the body\\ \midrule
\multirow{5}{*}{\textbf{A61J}}
& Containers specially adapted for medical or pharmaceutical purposes\\
& Devices or methods specially adapted for bringing pharmaceutical products into particular physical or administering forms\\
& Devices for administering food or medicines orally\\
& Baby comforters\\
& Devices for receiving spittle\\ \midrule
\textbf{A61K} & Preparations for medical, dental, or toilet purposes\\ \midrule
\multirow{4}{*}{\textbf{A61L}}
& Methods or apparatus for sterilising materials or objects in general\\
& Disinfection, sterilisation, or deodorisation of air\\
& Chemical aspects of bandages, dressings, absorbent pads, or surgical articles\\
& Materials for bandages, dressings, absorbent pads, or surgical articles\\ \midrule
\multirow{3}{*}{\textbf{A61M}}
& Devices for introducing media into, or onto, the body\\
& Devices for transducing body media or for taking media from the body\\
& Devices for producing or ending sleep or stupor\\ \midrule
\multirow{4}{*}{\textbf{A61N} }
& Electrotherapy\\
& Magnetotherapy\\
& Radiation therapy \\
& Ultrasound therapy \\ \midrule
\textbf{A61Q} & Specific use of cosmetics or similar toilet preparations \\ \bottomrule
\end{tabular}
\end{footnotesize}
\label{table:cpc_table}
\end{table}

\begin{figure}[!ht]
\centering
\includegraphics[width=0.7\textwidth]{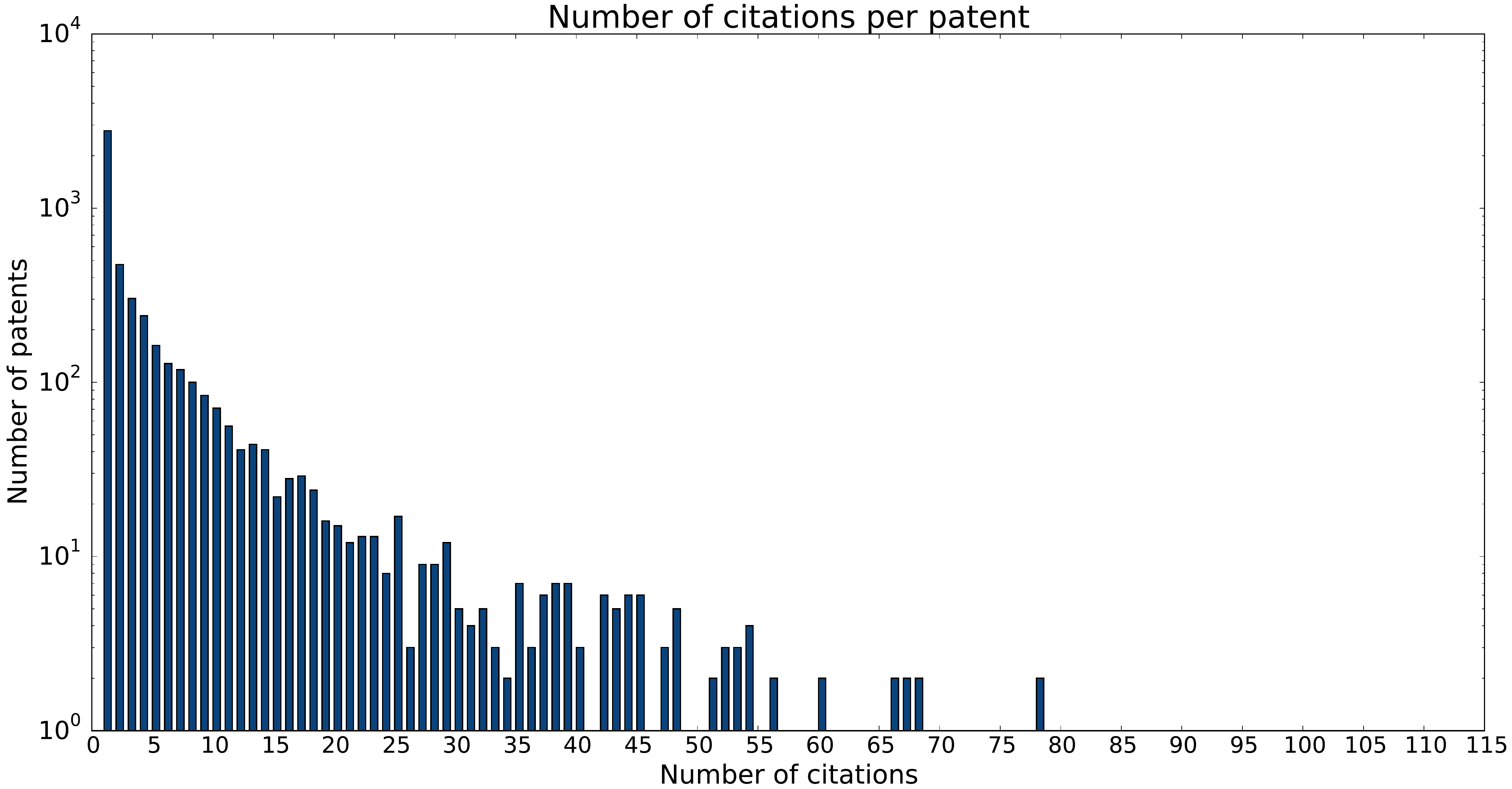}
\caption{\textbf{Number of patents on a logarithmic scale per number of citations for a subsample of 5000 randomly selected patents.}}\label{fig8}
\end{figure}

Finally, to get some insights into the existing search for prior art, we examine the distribution of the number of citations in the patent dataset. The citation counts for a subsample of 5000 randomly selected patents show that the distribution follows Zipf's law with many patents having very few citations and a low number of patents having many citations (Fig~\ref{fig8}).

\subsubsection*{Structure of a patent}
The requirements regarding the structure of a patent application are very strict and prescribe the presence of certain sections and what their content should be. For the automated comparison of texts it can be interesting to have a closer look at the different sections of the documents as it might, for instance, be sufficient to only compare a specific section of the texts. This can on the one hand be useful to perform a preliminary search for prior art before the patent text is written in its entirety in order to prevent unnecessary work and on the other hand, it can help to decrease the computational burden of preprocessing and comparing full texts.

The \textit{Patent Cooperation Treaty} (PCT) by the \textit{World Intellectual Property Organization} (WIPO) defines several obligatory sections a patent application must contain.\footnote{The WIPO is an agency of the \textit{United Nations} with the aim of unifying and fostering the protection of intellectual property.}
According to their requirements, a patent application should consist of a title, an abstract, the claims, and the description, where the invention is thoroughly described and the figures included in the document are explained in depth. Similar to scientific publications, a patent's abstract consists of a short summary of what the invention is about. The claims section plays a very special role in a patent application, as it defines the extent of the protection the patent should guarantee for the invention and is therefore the section the patent attorneys and patent officers base their search for prior art on. If the claims enter in conflict with already existing publications, they can be edited by weakening the protection requirements, which is why this section is reformulated the most during the possibly multiple stages of a patent process.

As both the \textsc{USTPO} and the \textit{European Patent Office} (\textsc{EPO})  adopt the PCT, the required sections are the same in the United States and in Europe. Nonetheless, some differences in the length of the description section can be observed. For a patent application handed in at the \textsc{USTPO}, this section mostly consists of the figures' descriptions, while for applications to the EPO it contains more abstract descriptions of the invention itself. This is due to stricter requirements of consistency between claims and description for European patents and must be taken into account when patents filed at different offices are compared, as this might result in lower similarity scores \cite{Piroi2017book4,Lupu2017book3}.

\subsection*{Constructing a labelled dataset with cited and random patents}
A first labelled dataset was constructed from the patent corpus by pairing up the patents and labelling each pair depending on whether or not one patent in the pair is cited by the other. More formally, let $\mathcal{P}$ be the set of patents in the corpus and $\mathcal{P}^2$ its Cartesian product. Each patent pair $(p_1, p_2) \in \mathcal{P}^2$ then gets assigned the label $1$ (\emph{cited}) if $p_2$ is contained in the search report of patent $p_1$ and $0$ (\emph{random}) otherwise. As some of the tested approaches are computationally expensive, we did not pair up all of the 100,000 documents in the corpus. Instead, the roughly 2,500 patents published in 2015 contained in the corpus were selected as a set of target patents and paired up with their respective citations as well as with a set of 1,000 randomly selected patents that were not contained in the search reports of any of the target patents.

Due to divisional applications and parallel filings and because claims are often changed during the application process, patents with the same description may appear several times with different IDs, which is why, as a sanity check, duplicates for some of the target patents were included in the dataset as well.\footnote{Duplicates are expected to receive a similarity score near or equal to 1.} All together, this `cited/random' labelled dataset consists of 2,470,736 patent pairs, of which 41,762 have a citation, 2,427,000 a random, and 1974 a duplicate relation.

\subsection*{Obtaining relevancy labels from a patent attorney}
As a subsample of the first dataset, our second dataset was constructed by taking ten of the target patents published in 2015, as well as their respective cited patents. In addition to that, in order to assess if relevant patents were missing from the search report, some of the random patents were included as well. These were selected based on their cosine similarity to the target patent, computed using the BOW vector representations. We chose for each patent the ten highest-scored, ten very low-ranked, and ten mid-ranked random patents. In total, this dataset subsample consists of 450 patent pairs, of which 151 are citations and 299 random pairs.

Neither knowing the similarity score of the patent pairs nor which ones were cited or random patents, the patent attorney manually assigned a score between 0 and 5 to the patent pairs according to how relevant the respective document was considered for the target patent, thus yielding the second labelled dataset. For most of the following evaluation, the patent attorney's scoring was transformed into a binary labelling by considering all patent pairs with a score greater than $2$ as relevant and the others as irrelevant.

\section{Supporting Information: Evaluation}

\subsection*{Computing AUC scores to evaluate similarity measures}\label{S3}
\begin{figure}[!ht]
\centering
\includegraphics[width=0.65\linewidth]{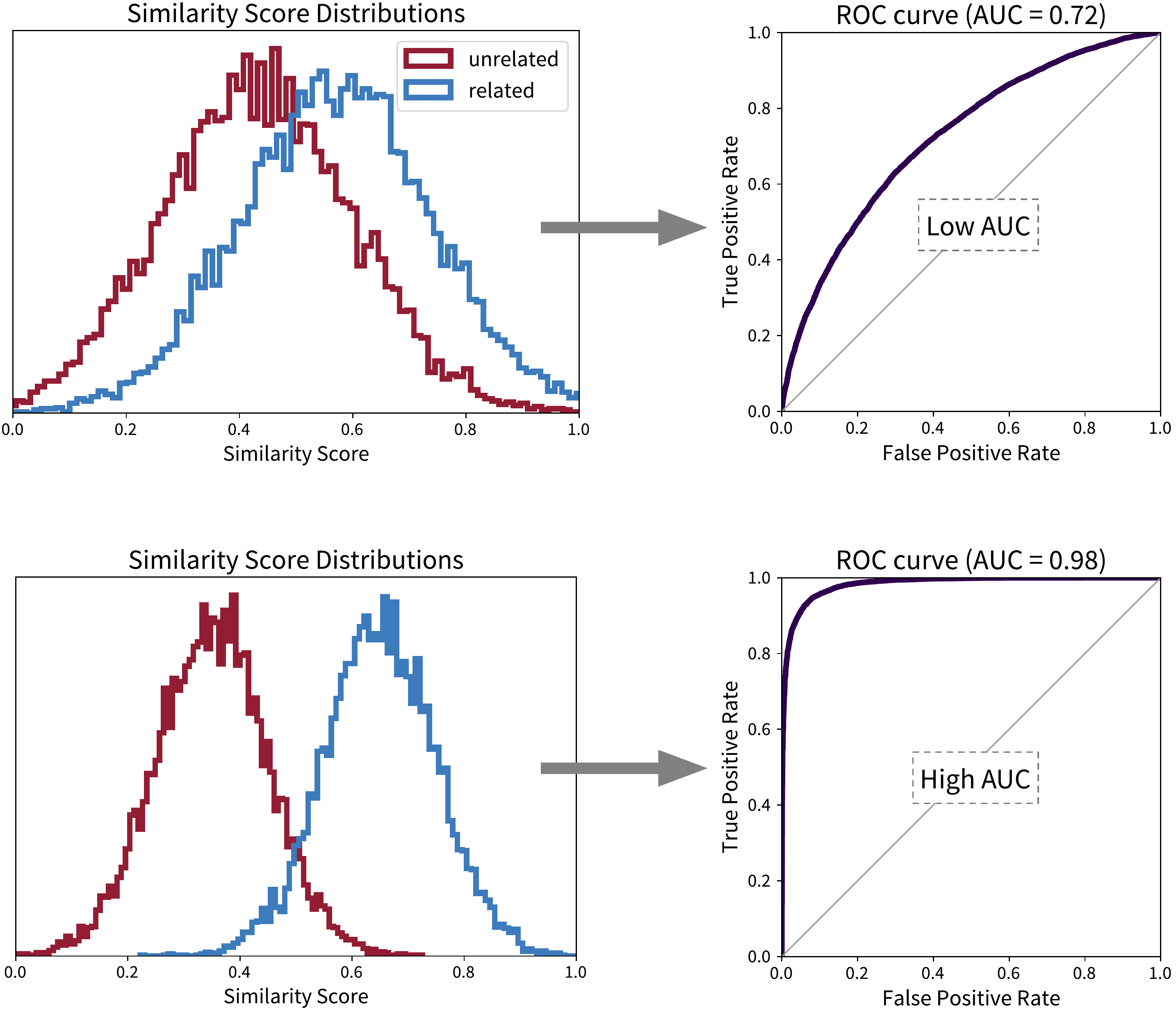}
\caption{ROC curve and AUC derived from similarity score distributions. Similarity scores were generated using artificial data to show how the difference in similarity scores for related \emph{(blue)} and unrelated sample pairs \emph{(red)} influence the ROC curves and therefore result in a lower or higher AUC.}\label{fig4}
\end{figure}

When computing similarity scores for all patent pairs, this results in two distributions of similarity scores, one for the positive samples (pairs of patents where one patent was cited by the other) and one for the negative samples (random patents). Ideally, these two distributions of scores would be separated, such that it is easy to chose a threshold to identify a positive or negative sample based on the corresponding similarity score of the patent pair (Fig~\ref{fig4}). To measure how well these two distributions are separated, we can compute the area under the \emph{receiver operating characteristic} (ROC) curve. Every possible threshold value chosen for separating positive from negative examples can lead to some pairs of unrelated patents to be mistakenly considered as relevant, what is called \emph{false positives} (FP), or to pairs of related patents mistakenly regarded as irrelevant, so-called \emph{false negatives} (FN). Correct decisions are either \textit{true negatives}~(TN), i.e., a pair of random patents that was correctly considered as irrelevant, or \textit{true positives}~(TP), which are correctly detected cited patents. Based on this, for every threshold value we can compute the \textit{true positive rate} (TPR), also called \emph{recall}, the \textit{false positive rate} (FPR), and the \textit{false negative rate} (FNR) to set wrong and correct decisions into relation:
\begin{align*}
    TPR &= \frac{TP}{TP + FN}, \\
    FPR &= \frac{FP}{FP + TN}, \\
    FNR &= \frac{FN}{TP + FN}.
\end{align*}
By plotting the TPR against the FPR for different decision similarity score thresholds, we then obtain the graph of the ROC curve, where the \textit{area under the \textsc{ROC} curve} (\textsc{AUC}) conveniently translates the performance of the similarity measure into a number between $0.5$ (no separation between distributions) and $1$ (clear distinction between positive and negative samples), as shown in Fig~\ref{fig4}.\footnote{Many information retrieval applications use precision and recall to measure the system's performance by comparing the number of relevant documents to the number of retrieved documents. However, since we do not only want to retrieve relevant documents, but in general select a discriminatory, interpretable, and meaningful similarity score, we consider the AUC, which relates the system's recall to its FPR.}

\section{Supporting Information: Results}

\subsection{Identifying cited patents using different similarity functions with BOW features}\label{S4a}
We evaluated all similarity measures listed in Table~\ref{table:simmeasures} using BOW features on the cited/random corpus. When computing the BOW features, we either used the term frequency (\emph{tf}) or a binary flag ($0/1$) for each word occurring in a document and experimented with raw values as well as values weighted by the words' \emph{idf} scores. Furthermore, these feature vectors were either normalized by the vector's maximum value or its length. The AUC scores for all these combinations can be found in Table~\ref{tab:auc_all_results}.
\begin{table}[!ht]
\centering
\caption{\textbf{AUC scores for all the tested combinations of BOW feature extraction approaches and similarity functions on the cited/random corpus.} The best result for each similarity function is printed in bold and the best result for each function class is underlined.\\
$^*$ The linear kernel with length normalized vectors corresponds to the cosine similarity.\\
$^+$ The AUC is equal, as for length normalized vectors (i.e.~$\Vert\mathbf{x}_i\Vert_2 = \Vert\mathbf{x}_j\Vert_2 = 1$), we get\\
$\Vert\mathbf{x}_i - \mathbf{x}_j\Vert_2^2 = (\mathbf{x}_i-\mathbf{x}_j)^T(\mathbf{x}_i-\mathbf{x}_j) = \mathbf{x}_i^T\mathbf{x}_i - 2 \mathbf{x}_i^T\mathbf{x}_j + \mathbf{x}_j^T\mathbf{x}_j = 1 - 2 \mathbf{x}_i^T\mathbf{x}_j  + 1 = 2-2 \mathbf{x}_i^T\mathbf{x}_j$\\
and $\mathbf{x}_i^T\mathbf{x}_j$ is equal to the cosine similarity.}
\begin{footnotesize}
\begin{tabular}{llcccc}
\toprule
\multicolumn{6}{c}{\textbf{Similarity coefficients}}\\
\midrule
      &    normalization    &  \emph{tf} &  0/1 &  \emph{tf-idf} &  0/1-\emph{idf} \\
\midrule
Braun-Blanquet & length &             0.8550 &               0.7941 &       \textbf{0.9480} &         0.8791 \\
      & max &             0.8075 &               0.7941 &       0.9338 &         0.8753 \\
Czekanowski, S{\o}rensen-Dice  & length &             0.8749 &               0.8371 &       \textbf{0.9555} &         0.9021 \\
      & max &             0.8593 &               0.8680 &       0.9505 &         0.9144 \\
Jaccard & length &             0.8749 &               0.8371 &       \textbf{0.9555} &         0.9021 \\
      & max &             0.8593 &               0.8680 &       0.9505 &         0.9144 \\
Kulczynski & length &             0.8767 &               0.8536 &       \textbf{\underline{0.9574}} &         0.9122 \\
      & max &             0.8761 &               0.9079 &       0.9571 &         0.9266 \\
Otsuka, Ochiai & length &             0.8759 &               0.8451 &       \textbf{0.9568} &         0.9072 \\
      & max &             0.8687 &               0.8982 &       0.9558 &         0.9323 \\
Simpson & length &             0.8566 &               0.8982 &       \textbf{0.9543} &         0.9268 \\
      & max &             0.8190 &               0.7879 &       0.9479 &         0.8685 \\
Sokal-Sneath, Anderberg & length &             0.8749 &               0.8371 &       \textbf{0.9555} &         0.9021 \\
      & max &             0.8593 &               0.8680 &       0.9505 &         0.9144 \\
\midrule
\multicolumn{6}{c}{\textbf{Kernel functions}}\\
\midrule
      &    normalization    &  \emph{tf} &  0/1 &  \emph{tf-idf} &  0/1-\emph{idf} \\
\midrule
Linear & length$^{*+}$ &             0.7336 &               0.8982 &       \textbf{\underline{0.9560}} &         0.9470 \\
      & max &             0.5411 &               0.7142 &       0.9387 &         0.8168 \\
Gaussian & length &             0.7336 &               0.8982 &       \textbf{0.9560} &         0.9470 \\
      & max &             0.6909 &               0.5010 &       0.6366 &         0.5083 \\
Histogram intersection & length &             0.7853 &               0.8050 &       \textbf{0.9239} &         0.8759 \\
      & max &             0.6939 &               0.7142 &       0.8969 &         0.7694 \\
Polynomial & length &             0.7336 &               0.8982 &       \textbf{0.9480} &         0.9468 \\
      & max &             0.5411 &               0.7142 &       0.9383 &         0.8168 \\
Sigmoidal & length &             0.7336 &               0.8982 &      \textbf{0.9560} &         0.9470 \\
      & max &             0.5411 &               0.5000 &       0.9387 &         0.7971 \\
\midrule
\multicolumn{6}{c}{\textbf{Distance functions}}\\
\midrule
      &    normalization    &  \emph{tf} &  0/1 &  \emph{tf-idf} &  0/1-\emph{idf} \\
\midrule
Canberra & length &             0.5253 &               0.5479 &       \textbf{0.6523} &         0.6184 \\
      & max &             0.5259 &               0.5937 &       0.6072 &         0.5438 \\
Chebyshev & length &             0.6252 &               0.5686 &       0.6056 &         0.5473 \\
      & max &            \textbf{0.6271} &               0.5000 &       0.6162 &         0.5006 \\
Hellinger & length &             \textbf{0.8746} &               0.6709 &       0.7213 &         0.6183 \\
      & max &             0.8064 &               0.5937 &       0.6559 &         0.5788 \\
Jensen-Shannon & length &             \textbf{0.8607} &               0.6699 &       0.7028 &         0.6173 \\
      & max &             0.7889 &               0.5937 &       0.6415 &         0.5787 \\
Manhattan & length &             \textbf{0.7987} &               0.6486 &       0.6437 &         0.6002 \\
      & max &             0.7239 &               0.5937 &       0.5997 &         0.5767 \\
Minkowski ($p=3$) & length &             0.6765 &               \textbf{0.7203} &       0.6934 &         0.5930 \\
      & max &             0.6606 &               0.5937 &       0.6616 &         0.5846 \\
Euclidean & length$^+$ &             0.7336 &               0.8982 &       \textbf{\underline{0.9560}} &         0.9470 \\
      & max &             0.6909 &               0.5937 &       0.6366 &         0.5794 \\
$\chi^2$ & length &             \textbf{0.8476} &               0.6686 &       0.7567 &         0.6134 \\
      & max &             0.7739 &               0.5937 &       0.6180 &         0.5543\\
\bottomrule
\end{tabular}
\end{footnotesize}
\label{tab:auc_all_results}
\end{table}

For all similarity functions (excluding the Minkowski distance) the best result is obtained when using either \textit{tf} (distance functions) or \textit{tf-idf} (kernel functions,  similarity coefficients, as well as Canberra and Euclidean distance) feature vectors. This shows that it is important to consider how often each term occurs in the documents instead of only encoding its presence or absence. Another observation that can be made is that the majority of the highest AUC scores is obtained on the \textit{tf-idf} feature vectors, which give a more accurate insight on how important each term actually is for the given document and reduce the importance of \textit{stop words}. Except for the Chebychev distance, the final normalization of the vectors should be performed using their lengths and not their maximum values. This might be due to the fact that the length normalization takes all the vector entries into account and not only the highest one, which makes it less sensitive to outliers, i.e.~extremely high values in the vector. With length normalized vectors as input, the linear kernel is equal to the cosine similarity and can thus be included into the group of similarity coefficients.

All in all, except for the Euclidean distance, which gives the same AUC as the cosine similarity using normalized vectors, the kernel functions and similarity coefficients yield much better results than the distance measures, which shows that it is more important to focus on words the texts have in common instead of calculating their distance in the vector space. Among similarity coefficients and kernel functions, the former function class gives slightly more robust results. Given that similarity coefficients are especially designed for sequence comparison by explicitly taking into account their subsequences' overlap, they seem to be the appropriate function class for measuring similarity between the BOW feature vectors.

The cosine similarity is widely used in information retrieval~\cite{Crocetti2015, Huang2008, Yates1999} and is well suited to distinguish between cited and random patents as it assigns lower scores to random than to cited patent pairs and, additionally, reliably detects duplicates by assigning them a score near or equal to $1$ (Fig~\ref{fig2}).

\subsection{Detailed examination of outliers in the citation process}\label{S4b}
For a better understanding of the disagreements between the cited/random labelling and the cosine similarity scores compared to the relevant/irrelevant labelling, we take a closer look at a FP yielded by the cosine similarity as well as a FP yielded by both, the cosine similarity and the cited/random labelling. In addition to that, in the main text we gave an example of a FN, i.e.~a relevant patent that was missed by the patent examiner, but would have been found by our automated approach, as it received a high similarity score.

\begin{figure}[!ht]
\centering
\includegraphics[width=0.49\textwidth]{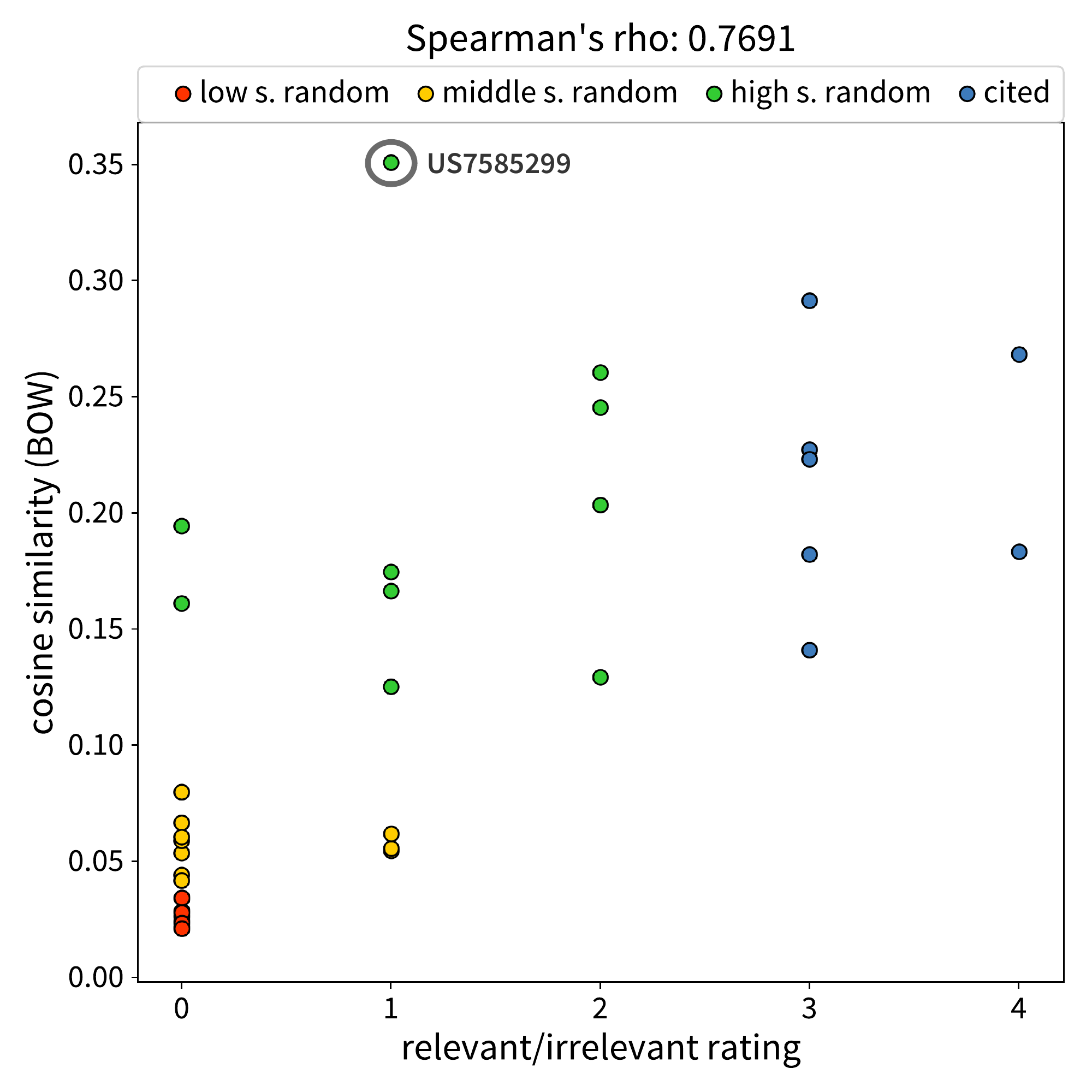}
\includegraphics[width=0.49\textwidth]{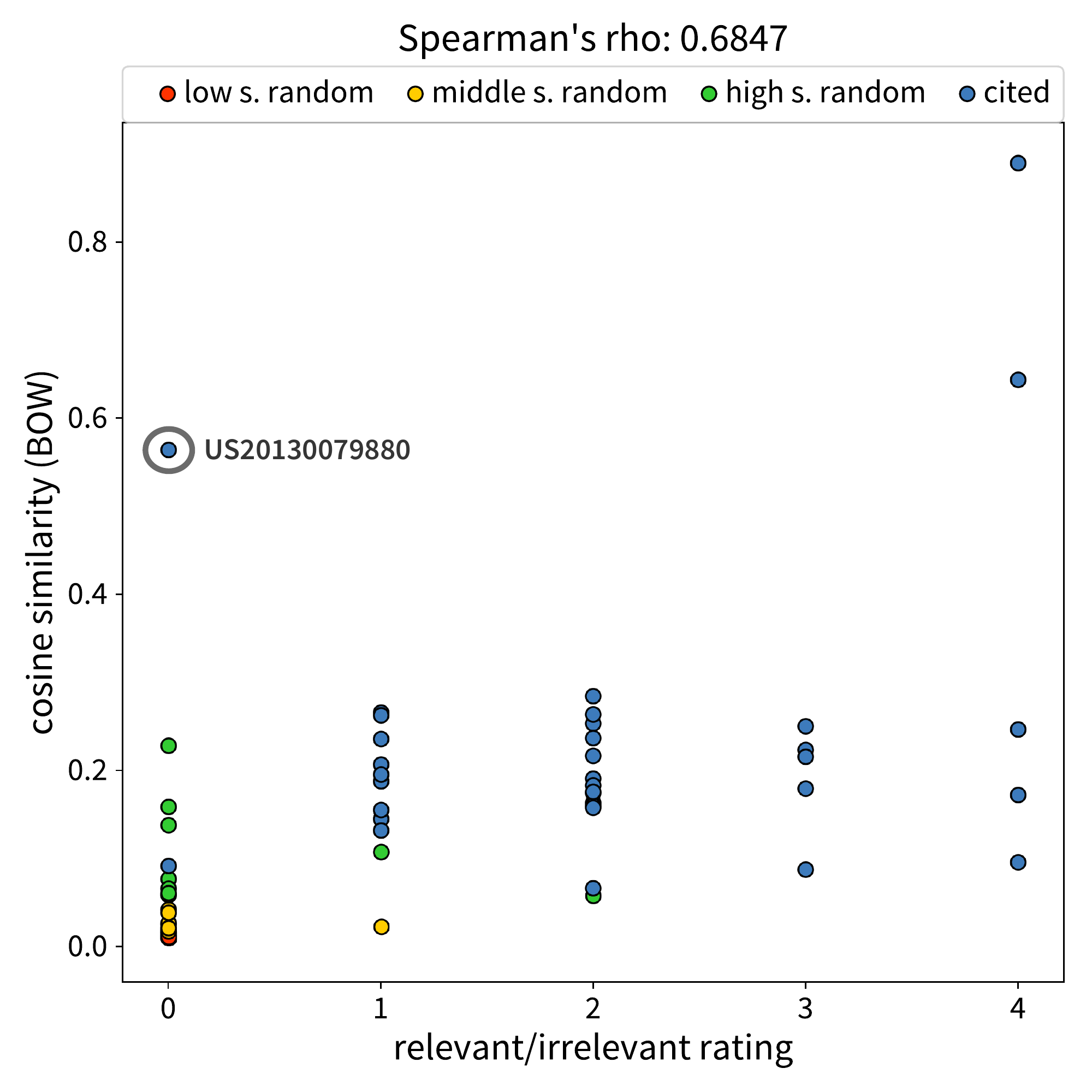}
\caption{\textbf{False positives.} \emph{Left:} Score correlation for the patent with ID US20150066086 with a false positive (ID~US7585299) yielded by the cosine similarity, circled in gray. \emph{Right:} Score correlation for the patent with ID US20150066087 with a false positive (ID~US20130079880) yielded by the cosine similarity and the cited/random labelling circled in gray. The blue dots correspond to the cited and the green, yellow, and red dots to the random patents whose colors describe whether they received a high, middle, or low cosine similarity score.}\label{fig9}
\end{figure}

\paragraph*{False positive yielded by our automated approach}
The patent with ID US7585299\footnote{\url{http://www.google.de/patents/US7585299}} marked with a gray circle in Fig~\ref{fig9} on the left would correspond to a FP taking both human labellings as the ground truth, because it received a high cosine similarity score although being neither relevant nor a citation.
The target patent (ID US20150066086\footnote{\url{http://www.google.de/patents/US20150066086}}) as well as the patent with ID~US7585299 describe inventions that stabilize vertebrae. In the target patent, the described device clamps adjacent spinous processes together by two plates held together by two screws without introducing screws inside the bones. The device described in patent US7585299, in contrast, stabilizes the spine using bone anchors, which are screwed e.g.~into the spinous processes or another part of the respective vertebrae and which have a clamp on the opposite end. The vocabulary in both patents is thus extremely similar, which leads to a high overlap on the BOW vector level, however, the two devices are far too different to be considered as similar inventions given that one is rigid and screwed into the bones whereas the other one only clamps the spinous processes and thereby guarantees a certain degree of flexibility.

\paragraph*{False positive yielded by our automated approach and the cited/random labelling}
For other target patents, more discordance with respect to the relevance of the other patents can be observed, also between the two human ratings.
The correlation of the relevant/irrelevant scoring for the patent with ID US20150066087\footnote{\url{http://www.google.de/patents/US20150066087}} in Fig~\ref{fig9} on the right shows that there are many cited patents that received a rather low score by the patent attorney, which means that the patent examiner produced a considerable amount of FP. One possible explanation for this might be that the patent examiners tend to produce rather more than less citations and thus include a large amount of the patents that are returned as results for their keyword query into the search report, although, on closer inspection, the relevance for the target patent is unfounded. This is also due to the fact that they mostly base their search on the claims section, which is usually kept as general as possible to guarantee a maximum degree of protection for the invention.
The analysis of the FP with ID~US20130079880\footnote{\url{http://www.google.de/patents/US20130079880}} (marked by the gray circle in the plot) underpins this hypothesis. The claims sections of the two patents are similar and the devices described in the patents are of similar construction, both having plates referred to as wings. The device described in the target patent, however, is designated to immobilize adjacent spinous processes, whereas the one described in patent US20130079880 is aimed at increasing the space between two adjacent vertebrae to relieve pressure caused for instance by dislocated discs.
Especially the similar claims section might have led the patent examiner to cite the patent, although the devices clearly have different purposes, which can easily be derived from their descriptions.

\end{document}